\documentclass[pre,twocolumn,showpacs]{revtex4}
\usepackage{epsfig}
\usepackage{amssymb}

\begin{document}

\title{The Fermi-Pasta-Ulam recurrence and related phenomena 
for 1D shallow-water waves in a finite basin}
\author{V. P. Ruban}
\email{ruban@itp.ac.ru}
\affiliation{Landau Institute for Theoretical Physics,
2 Kosygin Street, 119334 Moscow, Russia} 

\date{\today}

\begin{abstract}
In this work, different regimes of the Fermi-Pasta-Ulam (FPU) recurrence are simulated 
numerically for fully nonlinear ``one-dimensional'' potential water waves in a 
finite-depth flume between two vertical walls. In such systems, the FPU recurrence is
closely related to the dynamics of coherent structures approximately corresponding 
to solitons of the integrable Boussinesq system. A simplest periodic solution of the
Boussinesq model, describing a single soliton between the walls, is presented in an
analytical form in terms of  the elliptic Jacobi functions. In the numerical experiments, 
it is observed that depending on a number of solitons in the flume and their parameters,
the FPU recurrence can occur in a simple or complicated manner, or be practically absent.
For comparison, the nonlinear dynamics of potential water waves over nonuniform beds is 
simulated, with initial states taken in the form of several pairs of colliding solitons. 
With a mild-slope bed profile, a typical phenomenon in the course of evolution 
is appearance of relatively high (rogue) waves, while for random, relatively
short-correlated bed profiles it is either appearance of tall waves, 
or formation of sharp crests at moderate-height waves.
\end{abstract}


\maketitle

\section{Introduction}

Nearly integrable wave systems are known to exhibit the Fermi-Pasta-Ulam (FPU)
recurrence, when a (finite-size) system approximately repeats its initial state after 
some period of evolution. Starting from the first observation of this phenomenon in 
the famous numerical experiment \cite{FPU1955} with one-dimensional (1D) lattices of
nonlinear oscillators, the FPU recurrence and related phenomena were studied 
in many physical contexts (see, e.g., 
\cite{ZK1965,Z1973,T1977,YF1978,I1981,YL1982,AK1986,ZLY2003,OOSB1998,CL08,SEH01,Z05,BI05},
and references therein). In particular, Zabusky and Kruskal \cite{ZK1965} 
discovered the solitons with their highly nontrivial behaviour, when numerically
investigated a mechanism of the recurrence for spatially periodic solutions of the 
Korteweg-de-Vries (KdV) equation. Now the theory of solitons has developed into
one of the main branches of nonlinear science.

It is a well known fact that many integrable mathematical models have their origin 
in the theory of water waves. So, the two most famous integrable equations are the
KdV equation, first derived for weakly dispersive unidirectional shallow-water waves,
and the nonlinear Schroedinger equation (NLSE) which describes an envelope 
of a train of deep-water waves \cite{Z1968}. For deep-water waves, many analytical and
experimental results concerning the FPU recurrence are known
\cite{T1977,YF1978,I1981,YL1982,AK1986,ZLY2003,TW1999,SB2002,ChHw2007,Leblanc2009}.
As to the shallow-water regime, only some numerical studies for KdV 
and its higher-order generalizations were performed until recently 
(see, e.g., \cite{OOSB1998,CL08}), while the FPU phenomenon was never considered 
theoretically for long waves in a finite flume, and also it was never studied 
experimentally in the shallow-water regime. 

It is clear that KdV equation is not adequate for long waves in a finite basin 
where they reflect from the walls.
Fortunately, there is another integrable model, namely the Boussinesq system, 
which approximately describes bidirectional shallow-water waves and therefore it is
potentially useful for analytical study of the FPU recurrence in a finite-length flume 
(concerning integrability of the Boussinesq system, see
\cite{Kaup1975,Kupershmidt1985,Smirnov1986,ZhangLi2003}, and concerning deviations 
of water waves from exact integrability, see \cite{OOSB1998,CGHHS2006}).
However, at the moment we do not have yet a clear theory of FPU recurrence for the 
shallow water, based on the Boussinesq system. Perhaps, a future theory should
be built with the help of the sophisticated mathematical methods developed for
obtaining spatially periodic solutions of integrable systems
(in particular, for the Boussinesq model see \cite{Smirnov1986}). In the present work, 
such a general purpose is not achieved, though a family of periodic solutions is
derived here in an explicit analytical form through a simple ansatz, 
which corresponds to a single soliton periodically moving between the walls.
However, that solution is by no means the main result of our work; it just plays 
an auxiliary role, namely to provide nearly ''many-solitonic'' initial conditions
for highly accurate numerical experiments.
\begin{figure}
\begin{center}
   \epsfig{file=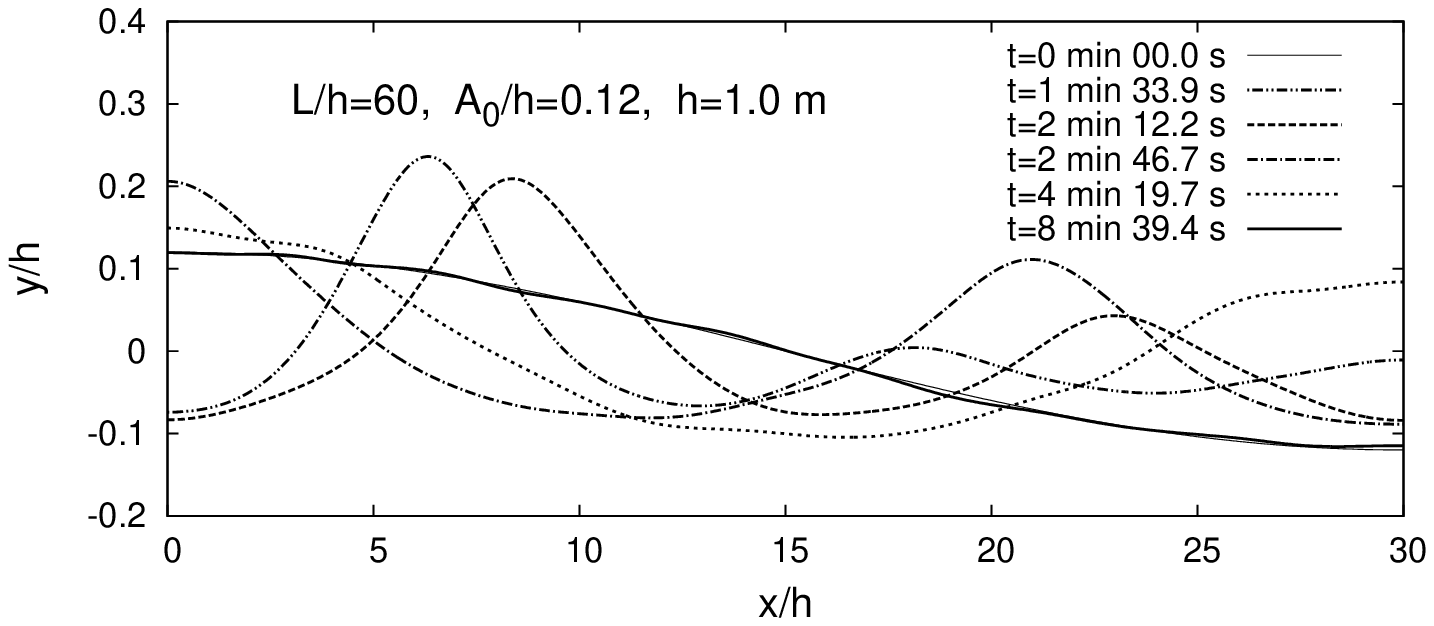,width=85mm}\\
   \epsfig{file=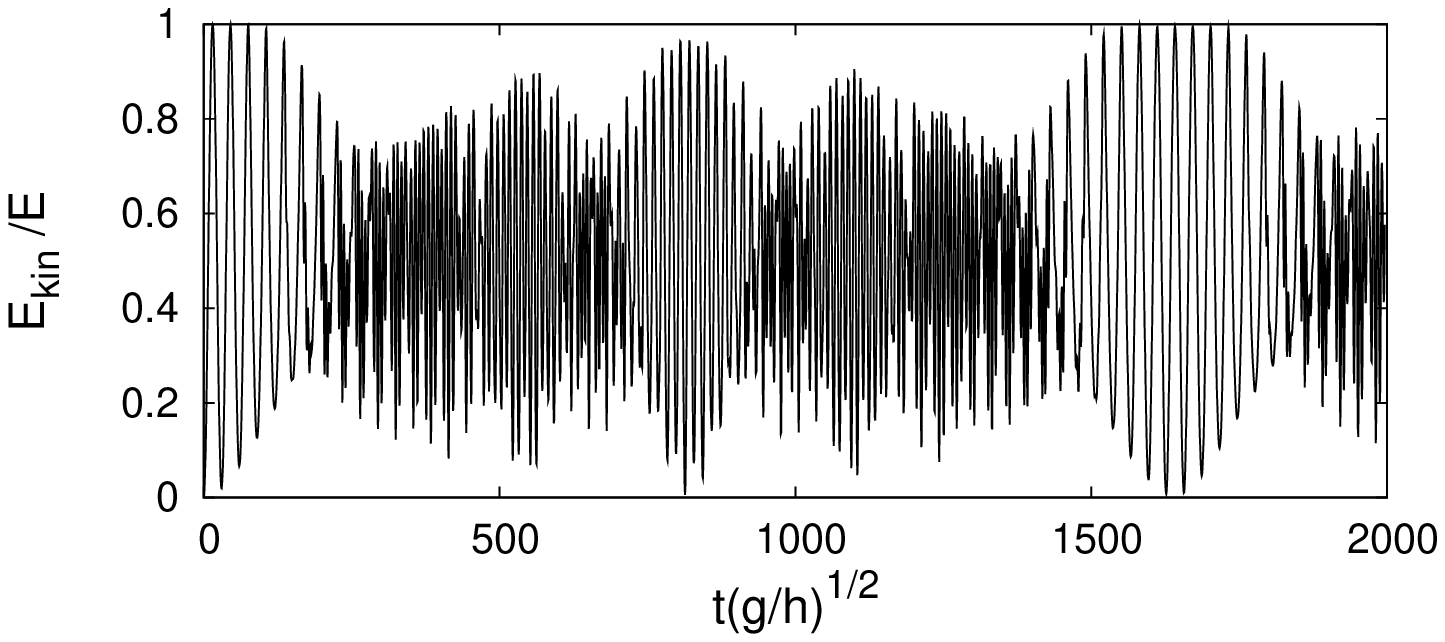,width=85mm}\\
   \epsfig{file=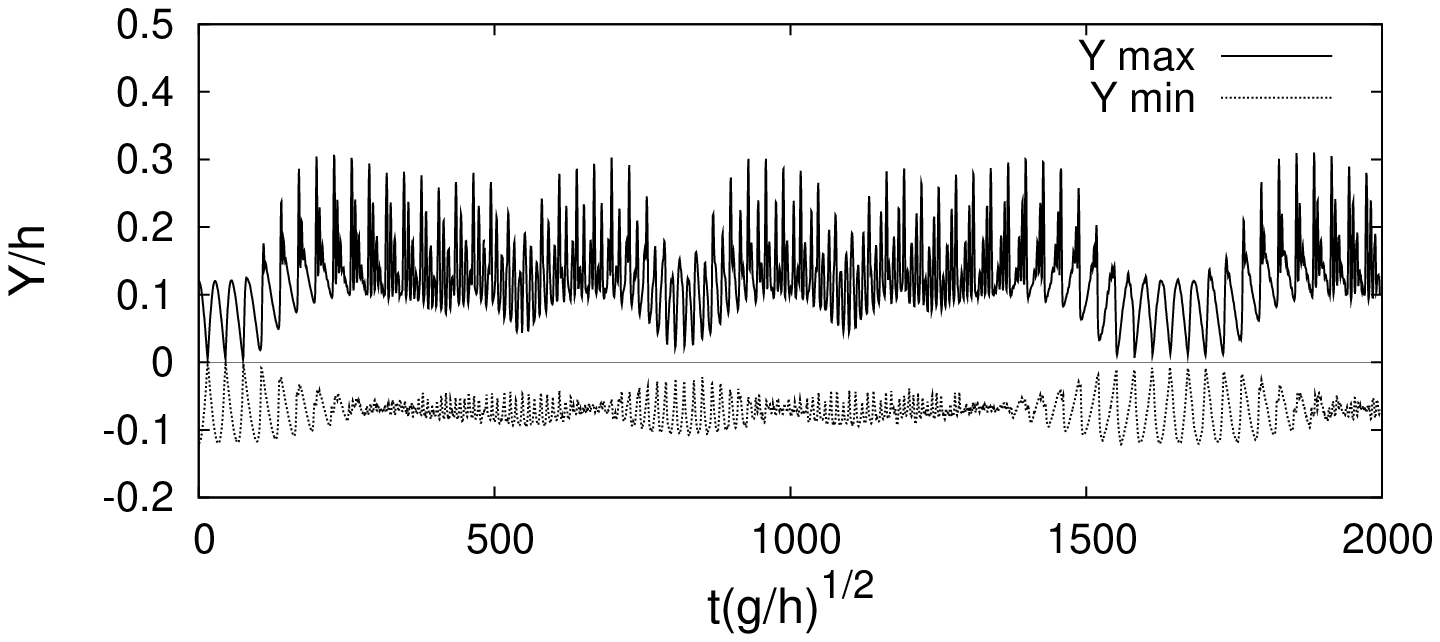,width=85mm}
\end{center}
\caption{The FPU recurrence is perfect with the initial shape of free surface in the form
$\eta_0(x)=0.12h\cos(2\pi x/60h)$:  
a) wave profiles at several time moments when the kinetic energy is at minimum;
b) the ratio of the kinetic energy to the total energy;
c) the maximum and minimum elevations of the free boundary.} 
\label{example_cos1A012} 
\end{figure}
\begin{figure}
\begin{center}
   \epsfig{file=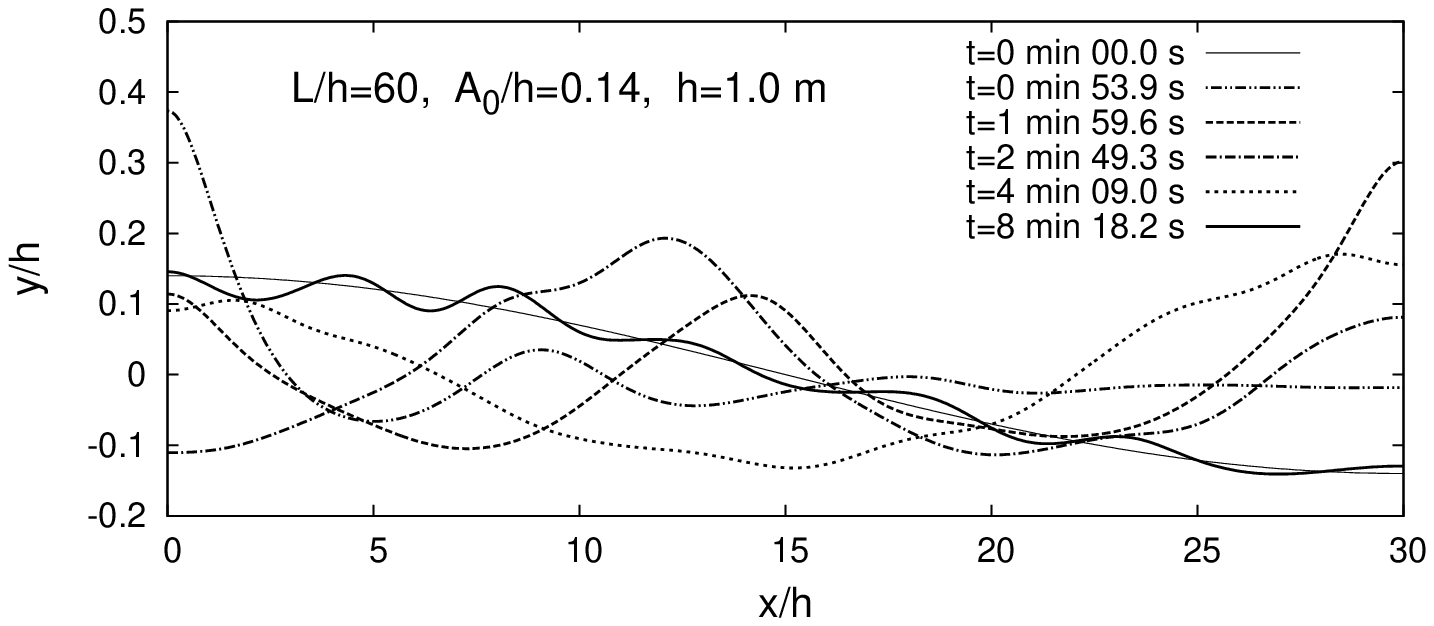,width=85mm}\\
   \epsfig{file=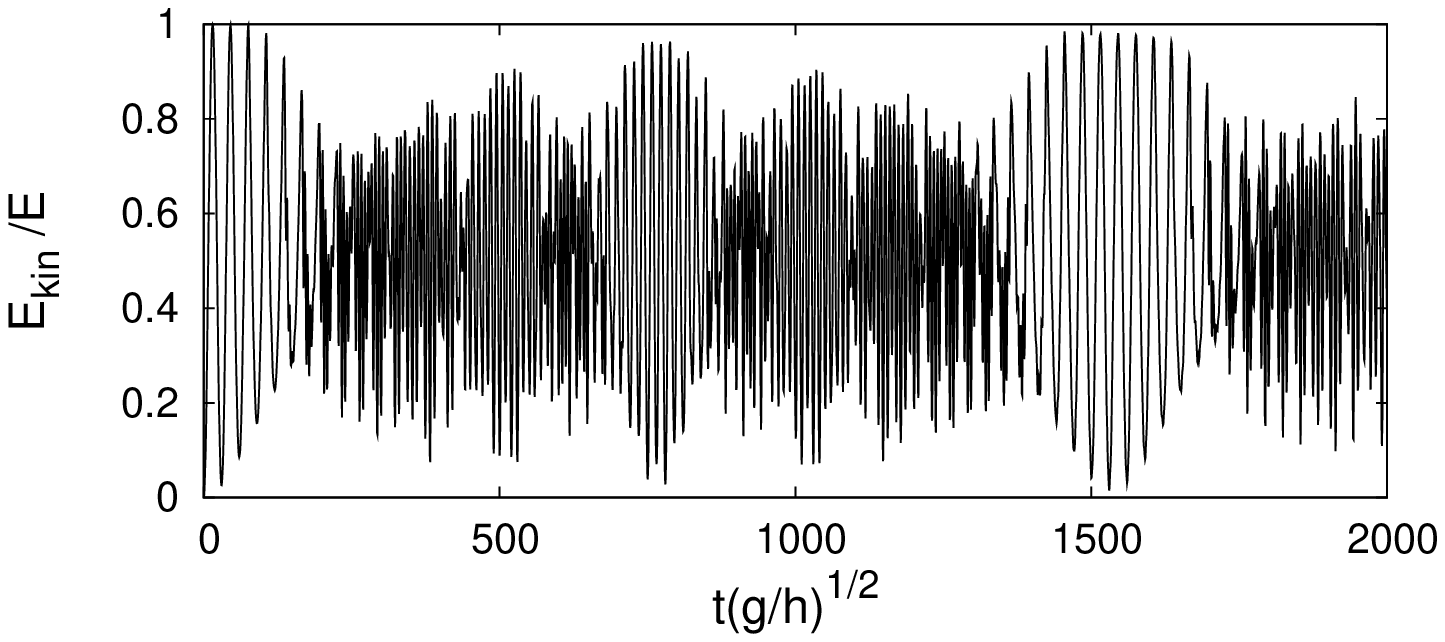,width=85mm}\\
   \epsfig{file=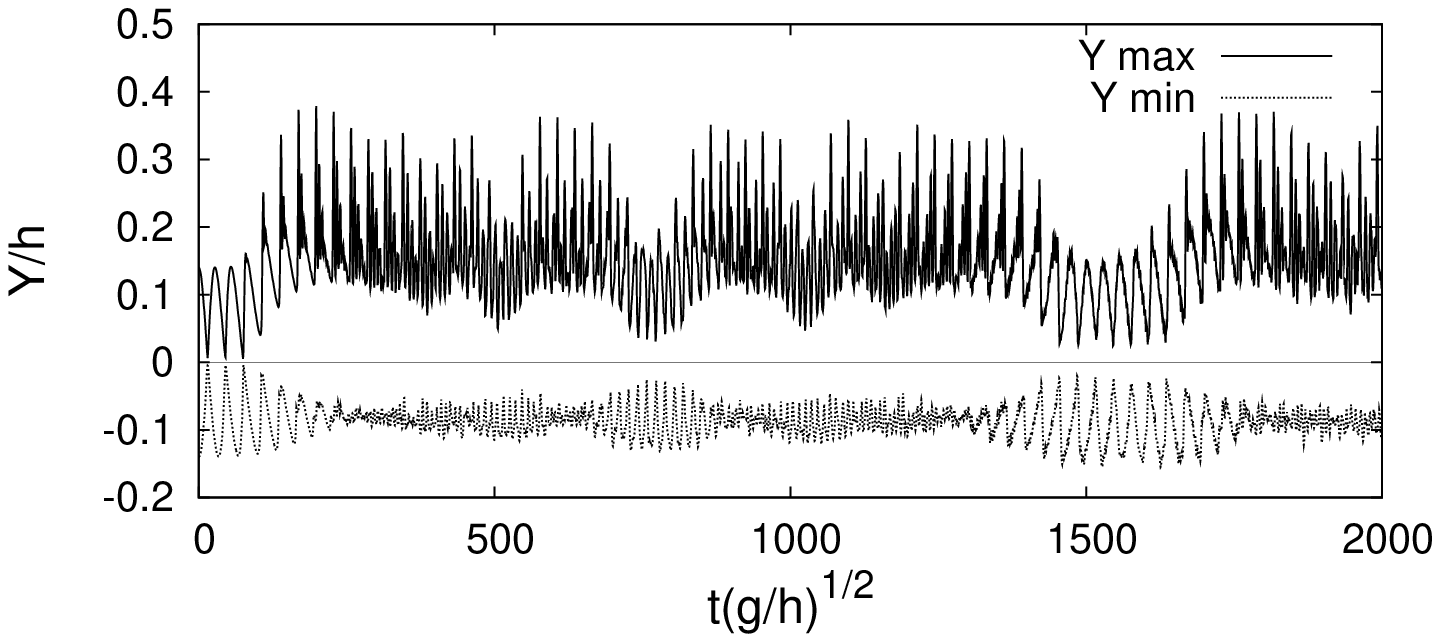,width=85mm}
\end{center}
\caption{The FPU recurrence is less perfect with larger initial amplitude, 
$\eta_0(x)=0.14h\cos(2\pi x/60h)$, because effects of non-integrability are stronger.} 
\label{example_cos1A014} 
\end{figure}

Very recently, a short paper by the present author was published, 
where for the first time the FPU recurrence was studied numerically for fully
nonlinear shallow-water waves in a finite flume  \cite{R2011Pisma}. 
Exact equations of motion for potential planar flows of a perfect fluid 
with a free surface in terms of so called conformal variables 
were employed in that study \cite{R2004PRE,R2008PRE}. The simplest initial states 
were taken, with zero velocity field and a cosine-shaped free boundary.
Two typical examples of recurrence are presented in 
Figs.\ref{example_cos1A012}-\ref{example_cos1A014}.

One of the purposes of the present work is to provide additional numerical examples 
of the recurrence for different initial states, and to demonstrate a relation of the 
FPU phenomenon in shallow-water finite basins to solitons of the approximate 
Boussinesq system. Another purpose is to observe what new effects appear in the dynamics
of long dispersive waves if the bottom boundary is nonuniform (it should be noted that 
conformal variables provide exact equations of motion for arbitrary nonuniform bottom 
profile when it is parametrized by an analytical function \cite{R2004PRE,R2008PRE}). 
In particular, three kinds of bed profiles will be considered, namely mild-slope beds, 
beds with quasi-random, relatively short-correlated corrugations, 
and beds with randomly placed barriers. 
The nonuniformity destroys the approximate integrability, so initial states in the form 
of several pairs of colliding solitons evolve to appearance of highly nonlinear wave events.
Such steep and tall waves can be considered as a 1D model for freak (rogue) waves 
sometimes arising in the coastal zone (the subject of freak waves is studied now very
extensively, see \cite{EJMBF2006,EPJST_185,NHESS2010}, and references therein).

\section{Different examples of the FPU recurrence}

\subsection{Notes about numerical method}

The numerical method employed here is based on the parametrization of ($x$-periodic) 
free surface in terms of a real function $\rho(\vartheta,t)$, 
as follows \cite{R2004PRE,R2008PRE}:
\begin{equation}
X+iY=Z(\vartheta + i\alpha(t)+\{1+i\hat{\mathsf R}_\alpha\}\rho(\vartheta,t)),
\end{equation}
where $\hat{\mathsf R}_\alpha$ is a linear integral operator diagonal in the 
discrete Fourier representation, ${\mathsf R}_\alpha(m)=i\tanh(\alpha m)$. Operator 
$\hat{\mathsf R}_\alpha$ depends parametrically on a positive quantity $\alpha(t)$  
(for details, see \cite{R2004PRE,R2008PRE}).
A fixed analytical function $Z(\zeta)$ determines conformal mapping of sufficiently 
wide horizontal stripe in the upper half-plane of an auxiliary complex variable $\zeta$,
adjacent to the real axis $\mbox{Im }\zeta =0$, onto a region in the physical 
$(x,y)$-plane, with the real axis $\mbox{Im }\zeta =0$ parametrizing the bed profile.

Since we consider purely potential planar flows of an ideal fluid, the velocity field
is completely determined by a real function $\psi(\vartheta,t)$, which is
the boundary value of the velocity potential at the free surface. 

Exact compact expressions for the time derivatives $\rho_t(\vartheta,t)$, 
$\psi_t(\vartheta,t)$, and $\dot\alpha(t)$ were obtained, 
corresponding to the dynamics of potential water waves 
in the uniform gravity field $g$ \cite{R2004PRE,R2008PRE}:
\begin{eqnarray}
&&\!\!\rho_t=-\mbox{Re}[\xi_\vartheta(\hat {\mathsf T}_\alpha+i){\mathsf Q}],
\label{rho_t_alpha} 
\\
&&\!\!\psi_t=-\mbox{Re}[\Phi_\vartheta(\hat{\mathsf T}_\alpha\! +i){\mathsf Q}] 
-\frac{|\Phi_\vartheta|^2}{2|Z'(\xi)\xi_\vartheta|^2}
-g\,\mbox{Im\,}Z(\xi),
\label{psi_t_alpha}
\\
&&\!\!\dot\alpha(t)=-\frac{1}{2\pi}\int_0^{2\pi}{\mathsf Q}(\vartheta)d\vartheta,
\label{dot_alpha}
\end{eqnarray}
where $\xi=\vartheta+i\alpha+(1+i\hat{\mathsf R}_\alpha)\rho$, 
$\Phi=(1+i\hat{\mathsf R}_\alpha)\psi$, and
${\mathsf Q}=(\hat{\mathsf R}_\alpha \psi_\vartheta)/|Z'(\xi)\xi_\vartheta|^2$.
Operator ${\mathsf T}_\alpha$ is diagonal in the discrete Fourier representation: 
${\mathsf T}_\alpha(m)=-i\coth(\alpha m)$ for $m\not=0$, and ${\mathsf T}_\alpha(0)=0$.
Eq.(\ref{rho_t_alpha}) is the so called kinematic boundary condition at the free surface,
written in terms of conformal variables, Eq.(\ref{psi_t_alpha}) is the dynamic
boundary condition (the Bernoulli equation), while Eq.(\ref{dot_alpha}) takes into account
temporal dependence of the conformal depth $\alpha$, which is necessary for conservation of
the total fluid volume.

If function $Z(\zeta)$ is expressed in terms of elementary analytical functions
[such as $\mbox{Exp}()$, $\mbox{Log}()$, and so on; for particular examples see 
\cite{R2004PRE,R2008PRE,R2005PLA,R2008PRE-2,R2008PRE-3}], then the right-hand sides of
Eqs.(\ref{rho_t_alpha})-(\ref{dot_alpha})
can be easily evaluated using the Fast Fourier Transform routines and 
mathematical library complex functions [so, in C programming language 
we have {\tt cexp()}, {\tt clog()}, and so on]. 
The above properties form the base of the numerical method.

In the numerical experiments, we use dimensionless variables 
(however, for graphical presentations the wave profiles are rescaled
to a characteristic depth $h=1$ m), and consider either flat horizontal, 
or $2\pi$-periodic nonuniform bed profiles [that means $Z(\zeta+2\pi)=2\pi+Z(\zeta)$] 
having an additional symmetry about the imaginary axis,
$\mbox{Im}Z(-\zeta'+i\zeta'')=\mbox{Im}Z(\zeta'+i\zeta'')$,
$\mbox{Re}Z(-\zeta'+i\zeta'')=-\mbox{Re}Z(\zeta'+i\zeta'')$.
This symmetry is required for simulations of waves between the vertical walls 
located at $x=0$ and at $x=\pi$. Of course, the functions $\psi(\vartheta,t)$ and
$\rho(\vartheta,t)$ should possess definite symmetries as well:
$\psi(\vartheta+2\pi,t)=\psi(\vartheta,t)$,
$\psi(-\vartheta,t)=\psi(\vartheta,t)$,
$\rho(\vartheta+2\pi,t)=\rho(\vartheta,t)$,
$\rho(-\vartheta,t)=-\rho(\vartheta,t)$.
The symmetries are automatically preserved in time if the initial data are symmetric.

In all our simulations, the system at $t=0$ is characterized by a free surface profile
$y=\eta_0(x)$, and by the velocity field ${\bf v}=0$. Such initial conditions with zero
kinetic energy $E_{\rm kin}$ were taken because they are convenient to observe the
recurrence by monitoring the time dependence of the quantity $E_{\rm kin}/E$, where $E$
is the total energy, which is conserved in the numerical experiments up to 7-8 decimal
digits. The function $\eta_0(x)$ is even and periodic, therefore it satisfies 
the boundary conditions $\eta'_0(0)=\eta'_0(L/2)=0$. 
A special procedure was designed, in order to find numerically a function 
$\rho(\vartheta,0)$ corresponding to a given initial profile $\eta_0(x)$ \cite{R2011Pisma}.

\begin{figure}
\begin{center}
   \epsfig{file=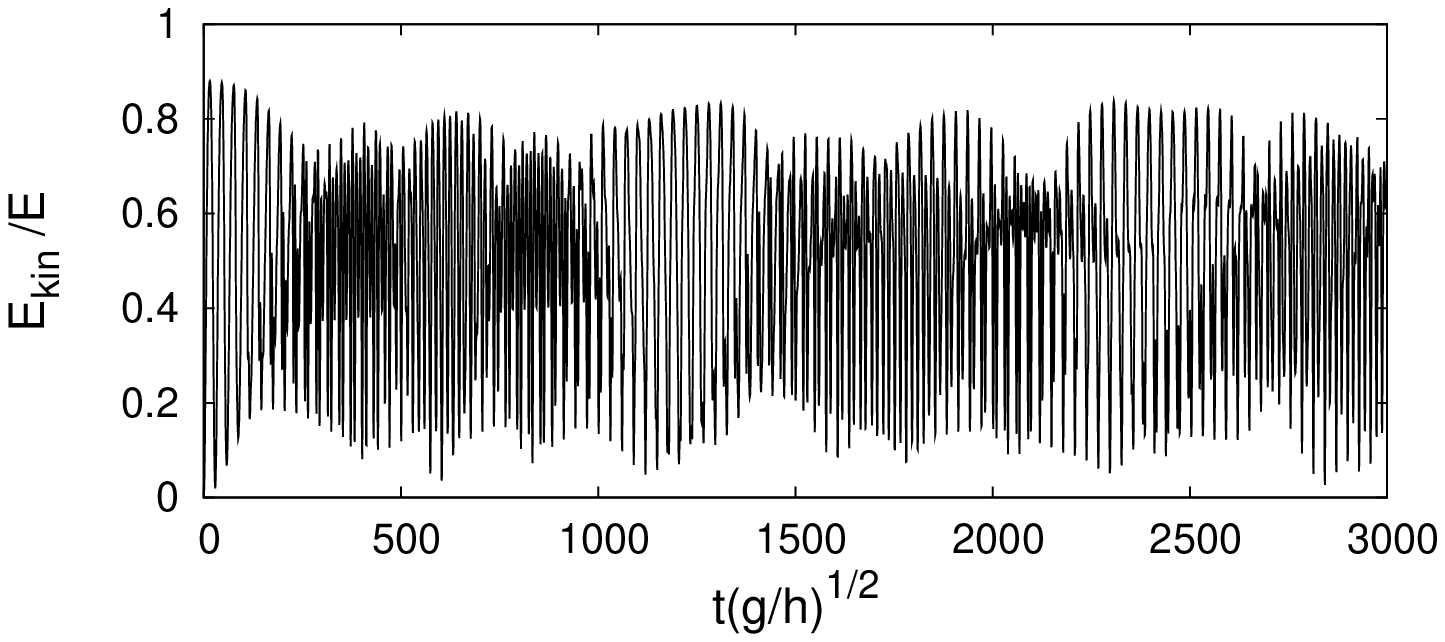,width=85mm}\\
   \epsfig{file=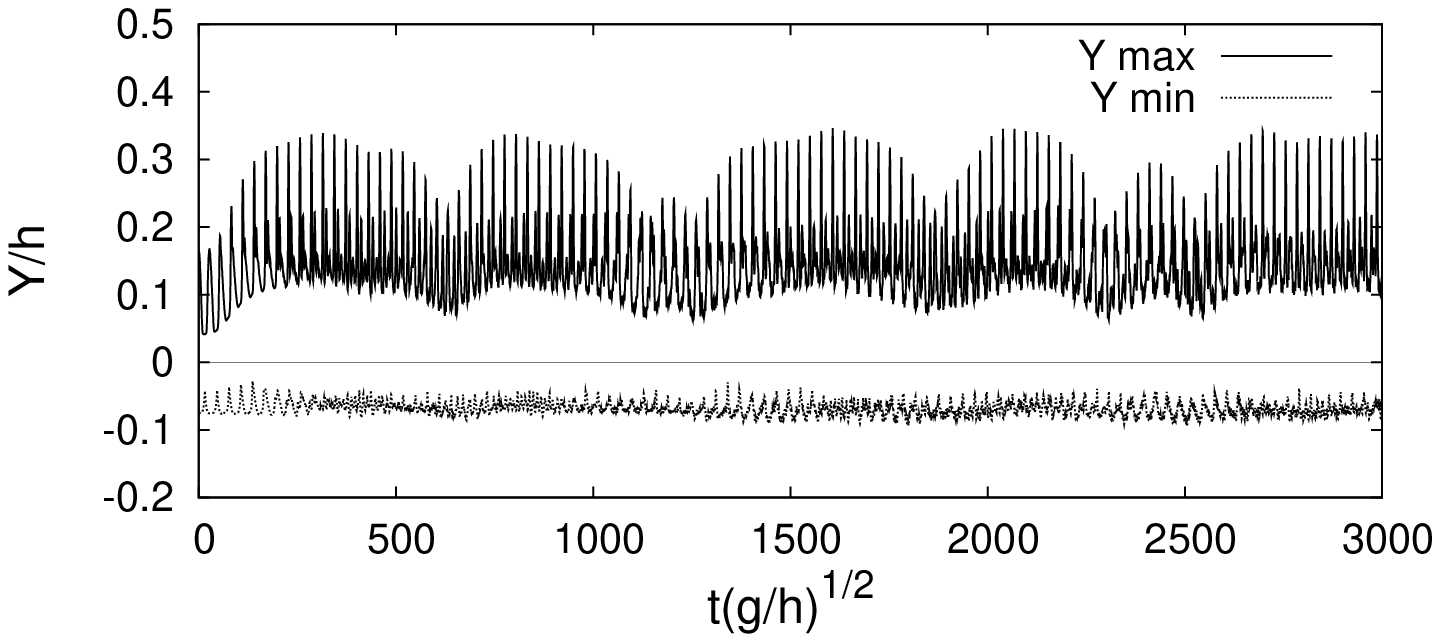,width=85mm}
\end{center}
\caption{The FPU recurrence is practically absent when 
$\eta_0(x)=0.24h\{[1+\cos(2\pi x/60h)]^3/8 -5/16\}$.} 
\label{example_cos3} 
\end{figure}

\begin{figure}
\begin{center}
   \epsfig{file=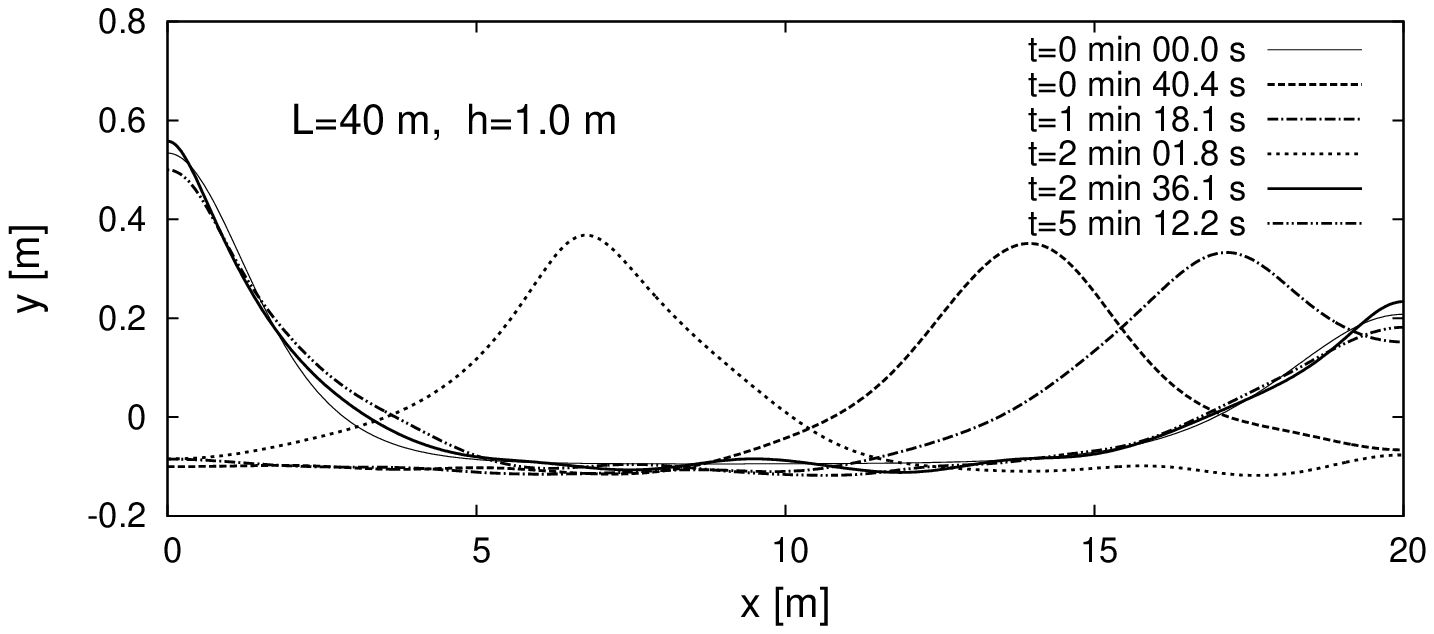,width=85mm}\\
   \epsfig{file=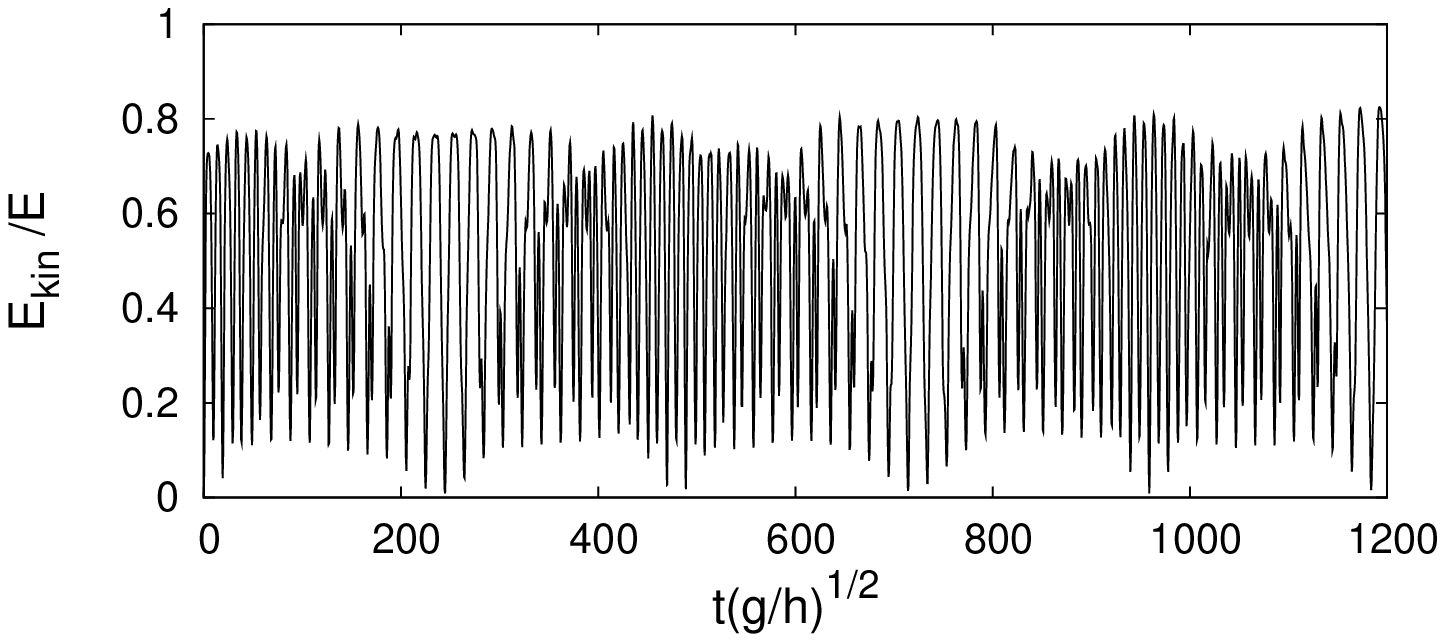,width=85mm}\\
   \epsfig{file=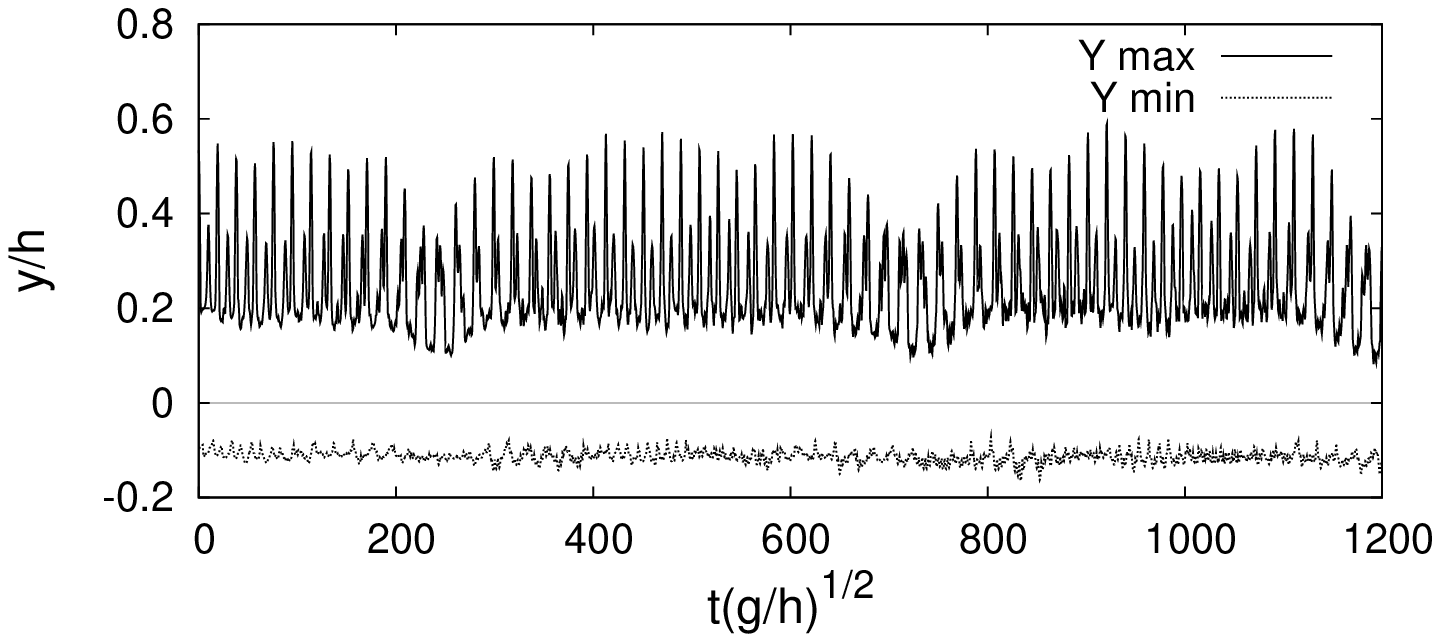,width=85mm}
\end{center}
\caption{Example of FPU recurrence when two solitons with different parameters 
are present in the system: 
a) wave profiles at several time moments when the kinetic energy is at minimum;
b) the ratio of the kinetic energy to the total energy;
c) the maximum and minimum elevations of the free boundary.} 
\label{example_A} 
\end{figure}

\subsection{Example when recurrence is absent}

It should be stressed that the FPU recurrence takes place for special 
initial conditions only. 
It is clear from the theoretical point of view that recurrence corresponds to a nearly
closed trajectory on a torus in a phase space of an integrable system. The dimensionality
of the torus is equal to the number of effectively excited degrees of freedom.
Typically, the frequencies of that motion are not rationally related.
Therefore, in the generic case recurrence is not observed.
As an example when the recurrence is practically absent, 
in Fig.\ref{example_cos3} we present evolution of some relevant parameters in the 
numerical experiment with $\eta_0(x)=0.24h\{[1+\cos(2\pi x/60h)]^3/8 -5/16\}$.

From this point of view, at the moment it looks somewhat miraculous that the initial
profiles in the simplest form $\eta_0(x)=A_0\cos(2\pi x/L)$ demonstrate rather perfect
recurrences despite the number of effectively excited degrees of freedom can be fairly
large, $N_{\rm s}= 5...7$ for $A_0/h=0.12$ and $L/h=100...120$, as in the numerical
experiments reported in \cite{R2011Pisma}.

\subsection{Recurrence in the dynamics of solitons}

\begin{figure}
\begin{center}
   \epsfig{file=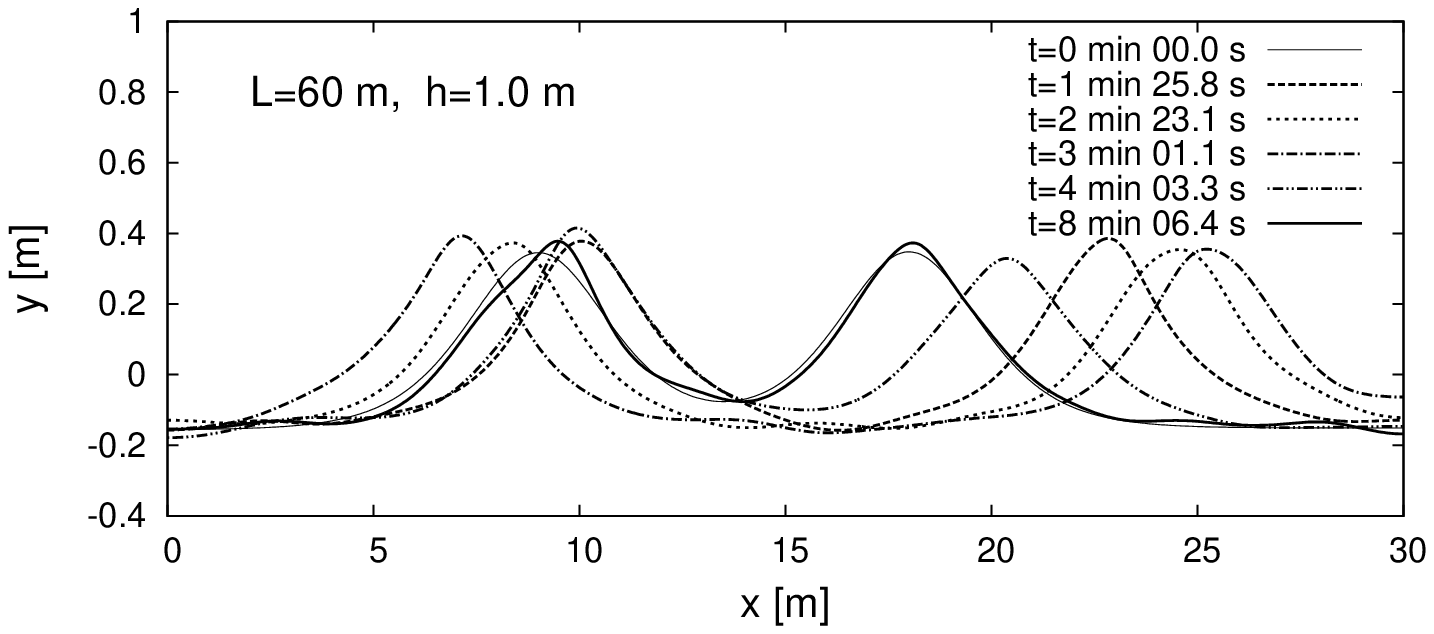,width=85mm}\\  
   \epsfig{file=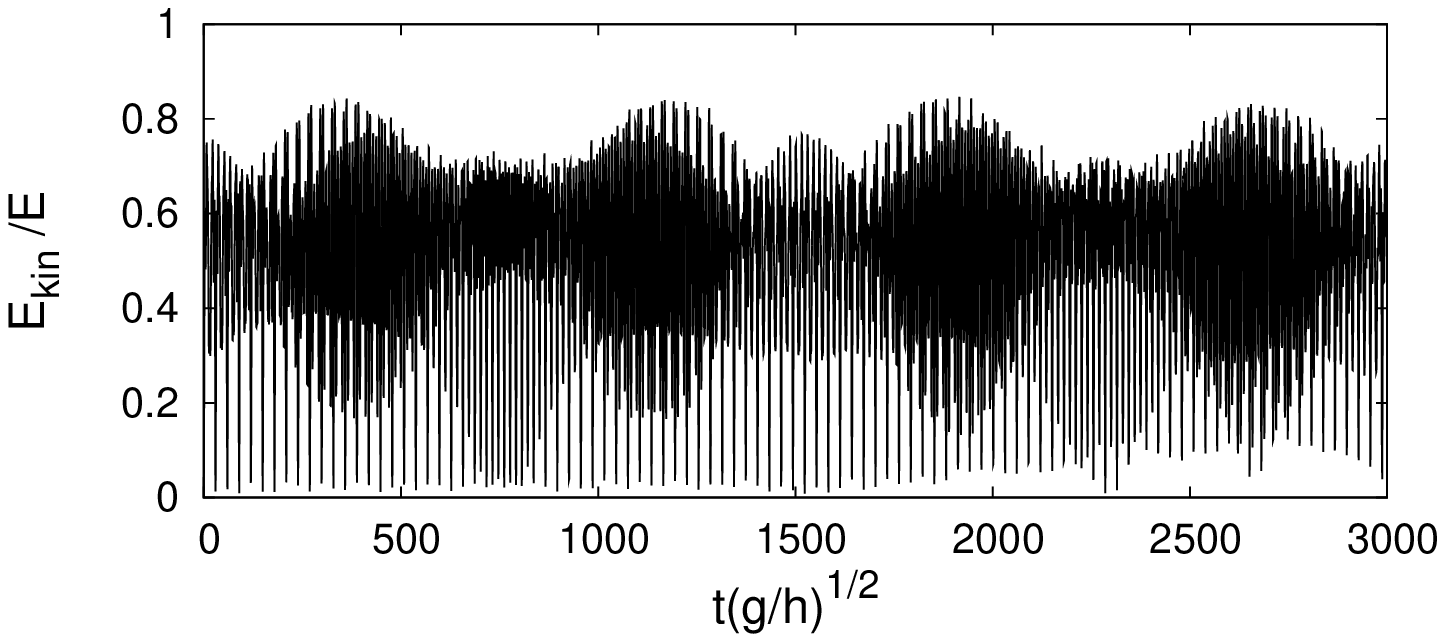,width=85mm}
\end{center}
\caption{Two pairs of solitons with equal parameters are present in the system: 
a) wave profiles at several time moments when the solitons are colliding;
b) the ratio of the kinetic energy to the total energy.} 
\label{example_B} 
\end{figure}

\begin{figure}
\begin{center}
   \epsfig{file=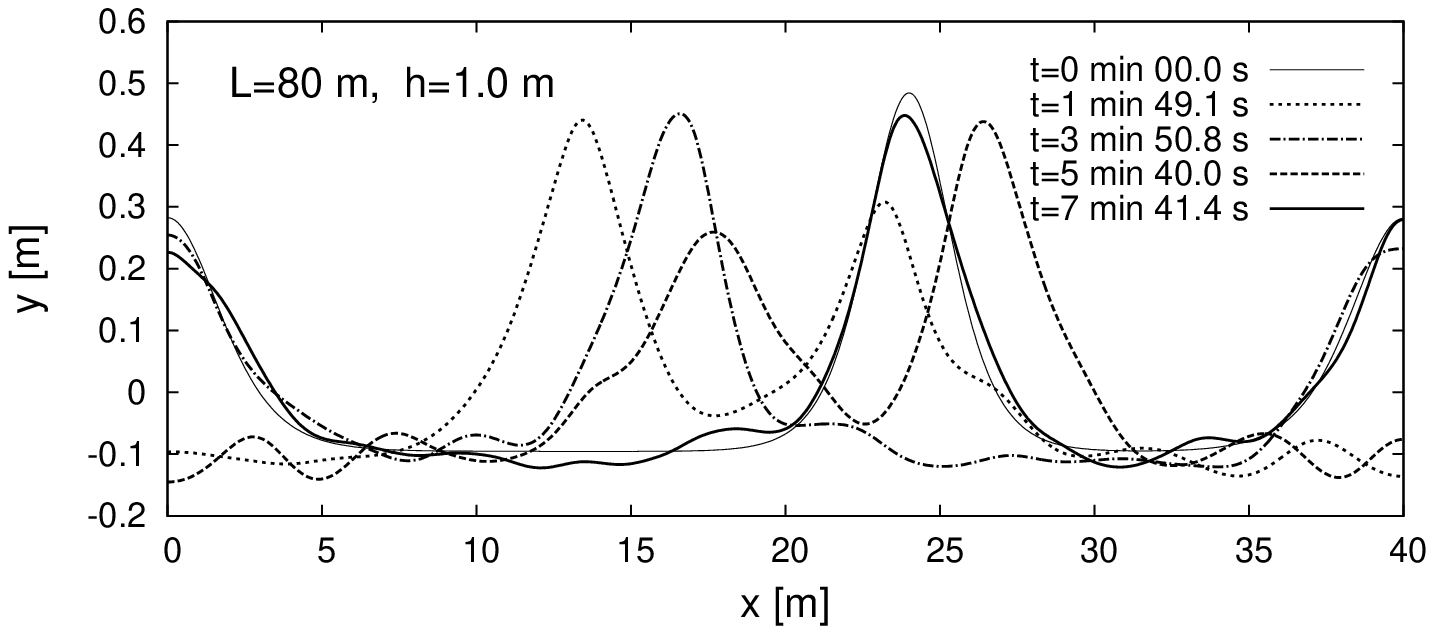,width=85mm}\\  
   \epsfig{file=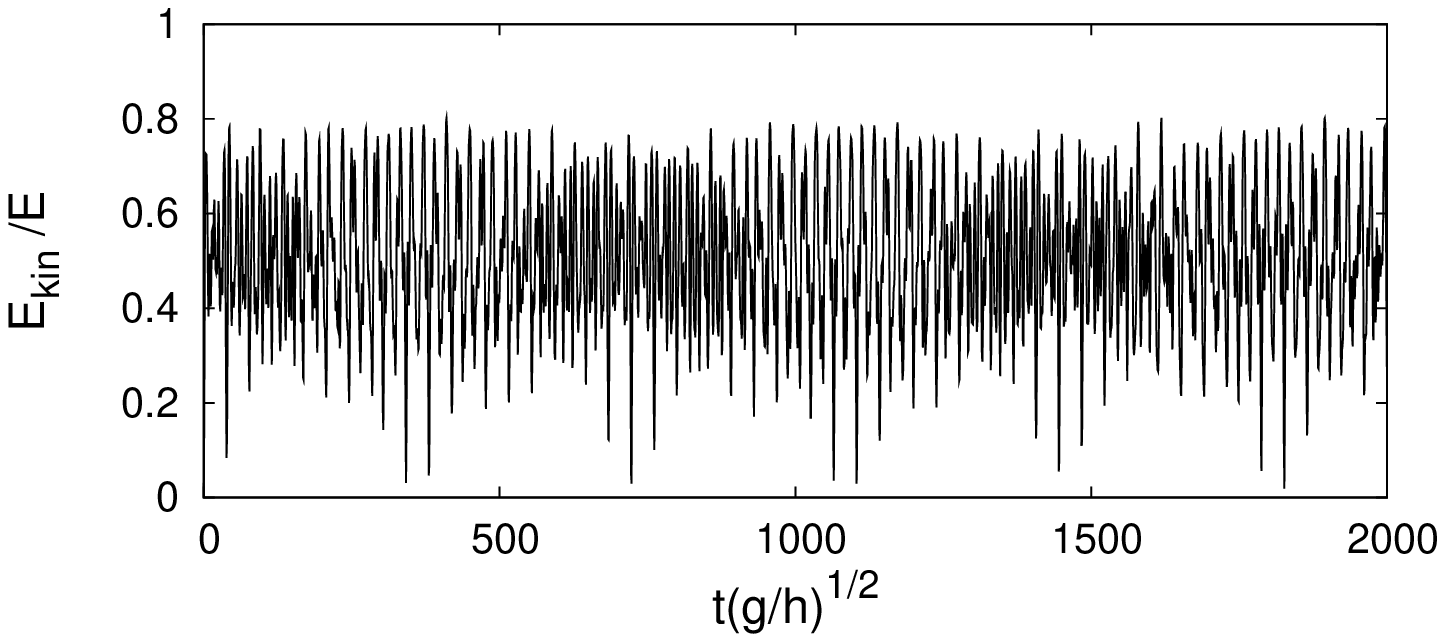,width=85mm}
\end{center}
\caption{Two pairs of solitons with different parameters are present in the system: 
a) wave profiles at several time moments when the solitons are colliding;
b) the ratio of the kinetic energy to the total energy.} 
\label{N_01A} 
\end{figure}

\begin{figure}
\begin{center}
   \epsfig{file=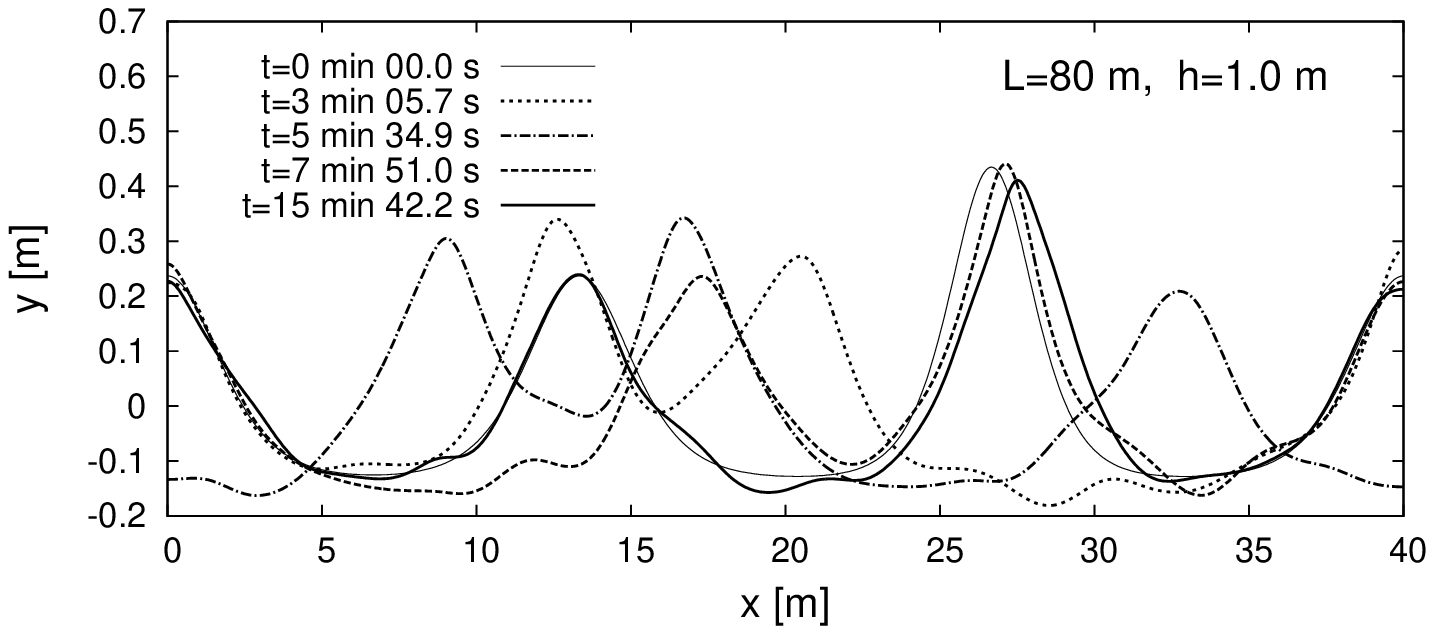,width=85mm}\\  
   \epsfig{file=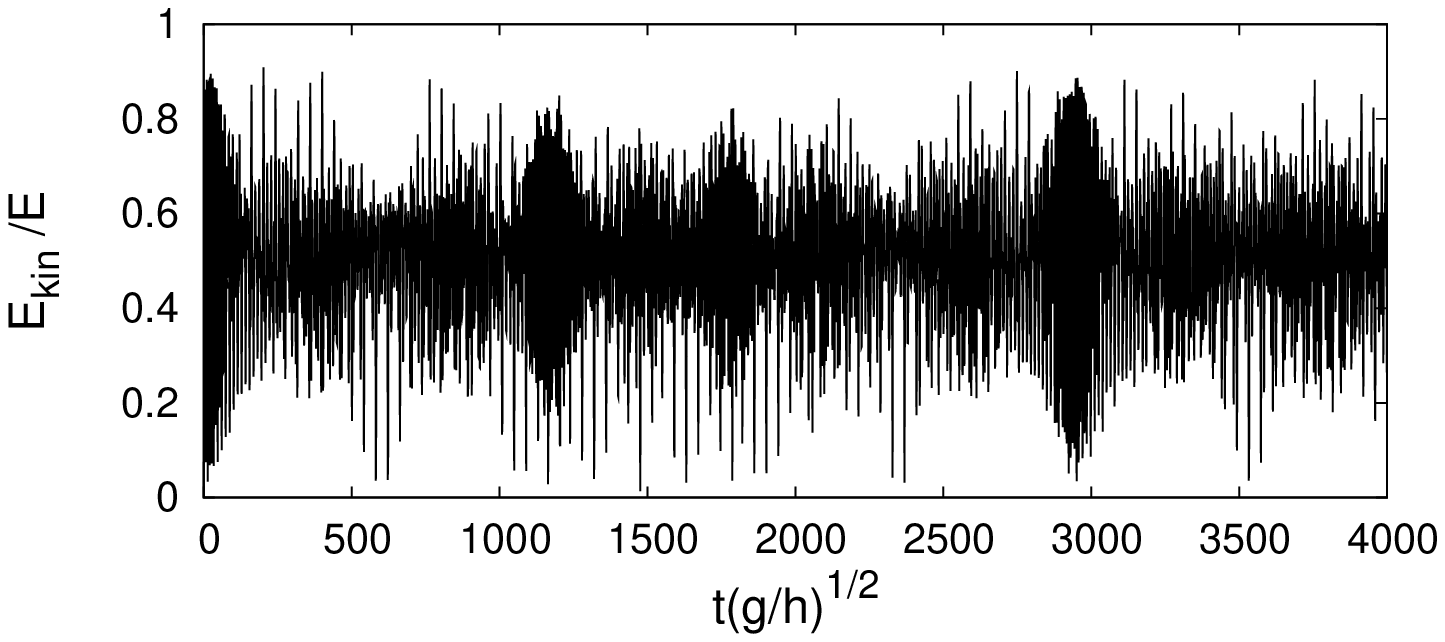,width=85mm}
\end{center}
\caption{More complicated regime of the FPU recurrence. Three pairs of solitons
are present, with one pair stronger than the two other.} 
\label{N_01D} 
\end{figure}

It is a well-known fact that integrable systems with periodic boundary conditions
posses so called finite-gap solutions, which are exactly finite-dimensional sub-systems
with the dynamics on a torus. For the Boussinesq system, the simplest example is given
in the Appendix, more involved cases are considered in \cite{Smirnov1986}. 
In general case, the formulas are quite complicated, and it is difficult to describe 
the corresponding degrees of freedom in terms of simple physical quantities. 
However, if the $x$-period is sufficiently long, an approximate description in terms of
several colliding solitons becomes possible. In the Boussinesq model, ``free''
solitons are characterized by positive or negative dimensionless velocities $s_n$ 
(constant, all different), and by their positions (phases) $x_n(t)$. 
The corresponding analytical solutions are presented in \cite{ZhangLi2003}. 
It is important that when two solitons with opposite velocities $s_1=-s_2=s$ collide 
at a position $x_0$ (or a single soliton collides with the wall), at some time moment
the velocity field is identically zero along the flow domain, while the shape of the
free surface is given by a simple formula $\eta(x)={\cal S}(x-x_0,s)$, with
\begin{equation}
\label{collision}
{\cal S}(x-x_0,s)=\frac{2h(s^2-1)}{\cosh^2[\sqrt{3(s^2-1)}(x-x_0)/2h]}.
\end{equation}
This formula was used in our numerical experiments to prepare initial states in the form
of several pairs of colliding solitons, placed sufficiently apart from each other.

In a finite domain, each soliton moves between the walls, and additionally it acquires
definite phase shift $\sigma(s_n)$ when it reflects from a wall ($s_n\to-s_n$
after reflection), and phase shifts
$\Delta(s_n,s_m)$  when it collides with other solitons \cite{ZhangLi2003}. 
In this picture, the recurrence takes place when at some time moment positions of all
solitons self-consistently return closely to their initial values. The simplest 
nontrivial example for quasi-recurrence in the system of two solitons is shown in
Fig.\ref{example_A}.

However, the above approximate description does not work if we initially put several 
identical humps, each corresponding to a pair of colliding solitons, 
to different positions. Fig.\ref{example_B} shows that the recurrence occurs 
in a more complicated way in such a case.

If we put two or more humps, with one of them higher than the others, then recurrence 
is possible only with tuned values of the larger velocity. Successful examples are
shown in Fig.\ref{N_01A} and Fig.\ref{N_01D}.

\begin{figure}
\begin{center}
   \epsfig{file=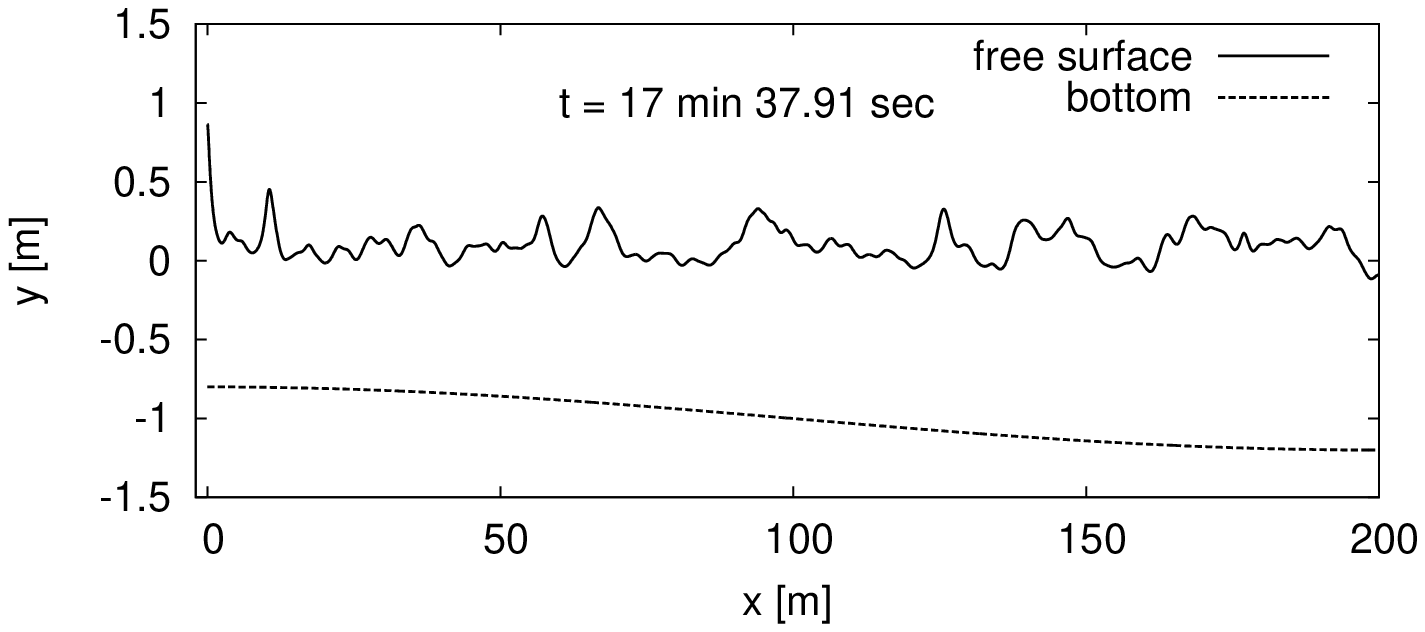,width=85mm}\\  
   \epsfig{file=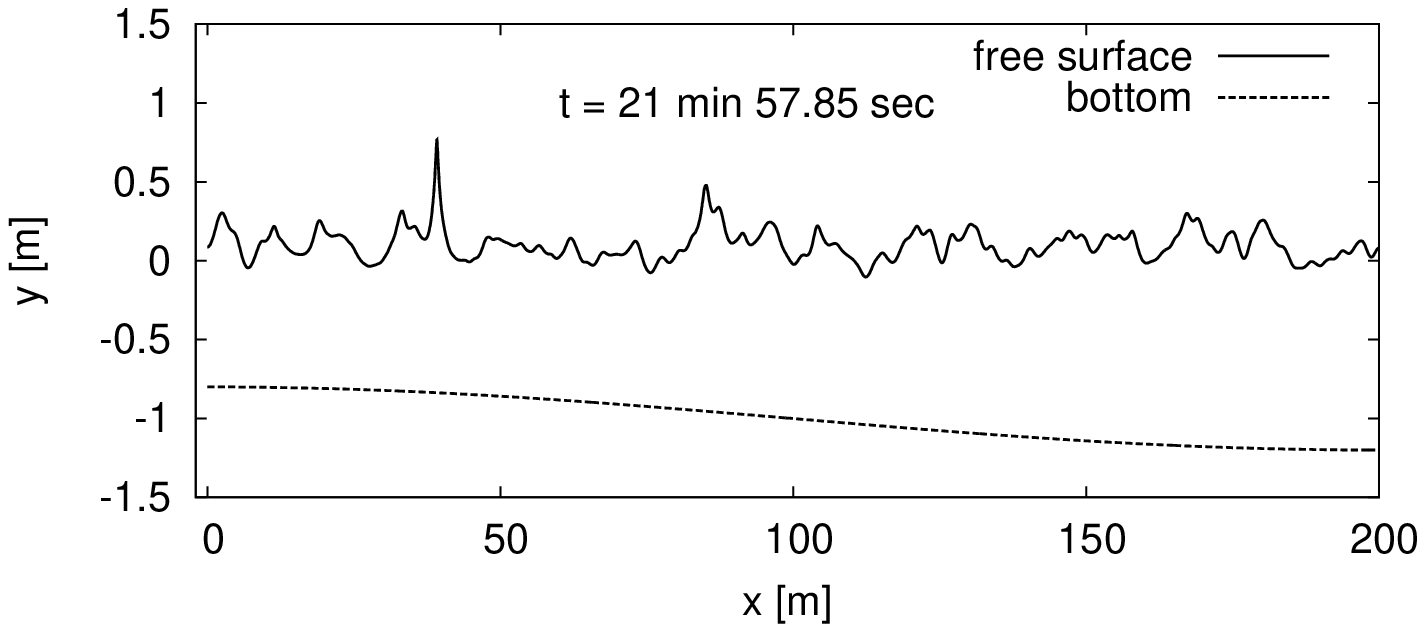,width=85mm}\\
   \epsfig{file=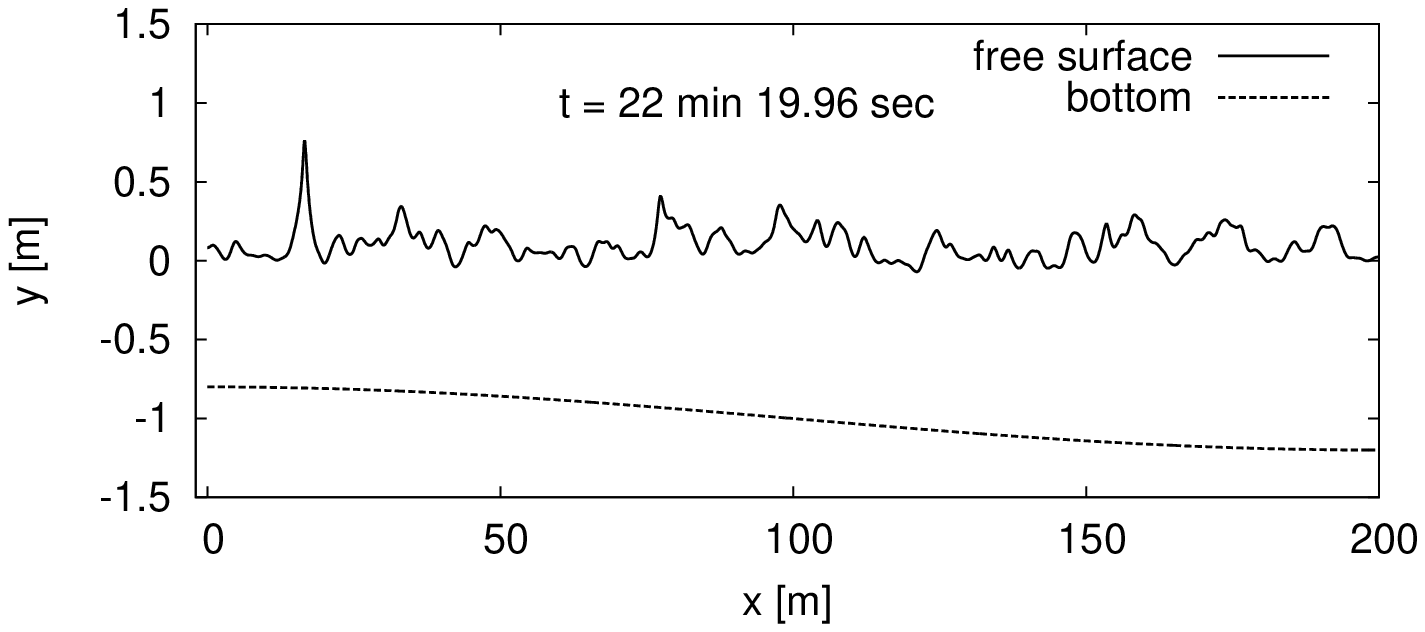,width=85mm}\\
   \epsfig{file=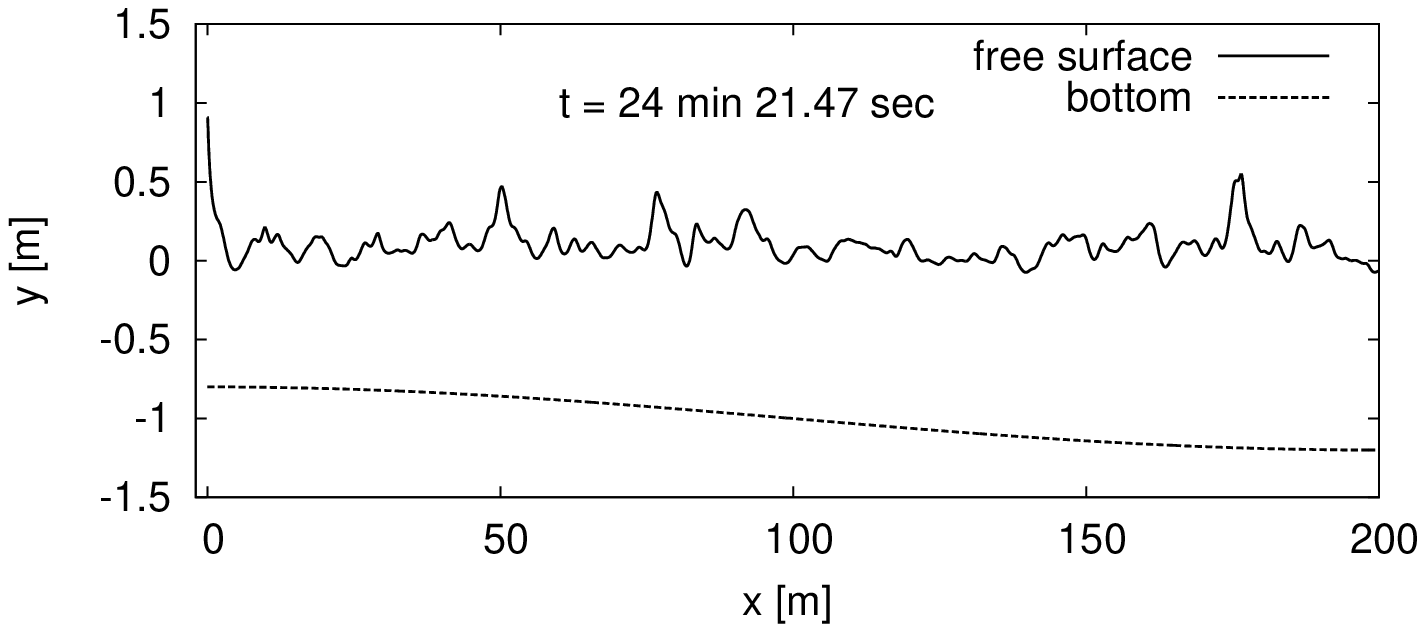,width=85mm}
\end{center}
\caption{Rogue waves appear over a mild-slope bed, some of them near the wall.} 
\label{smooth_bed_big_waves} 
\end{figure}

\begin{figure}
\begin{center}
   \epsfig{file=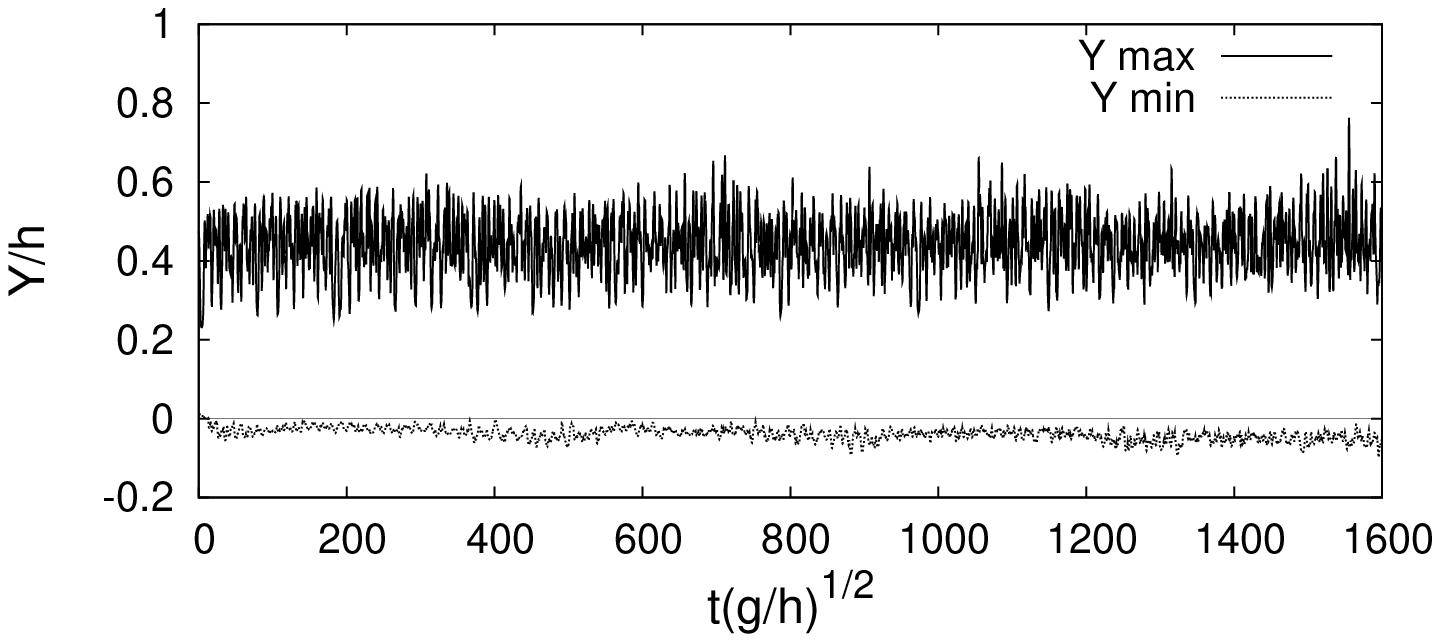,width=85mm}\\  
   \epsfig{file=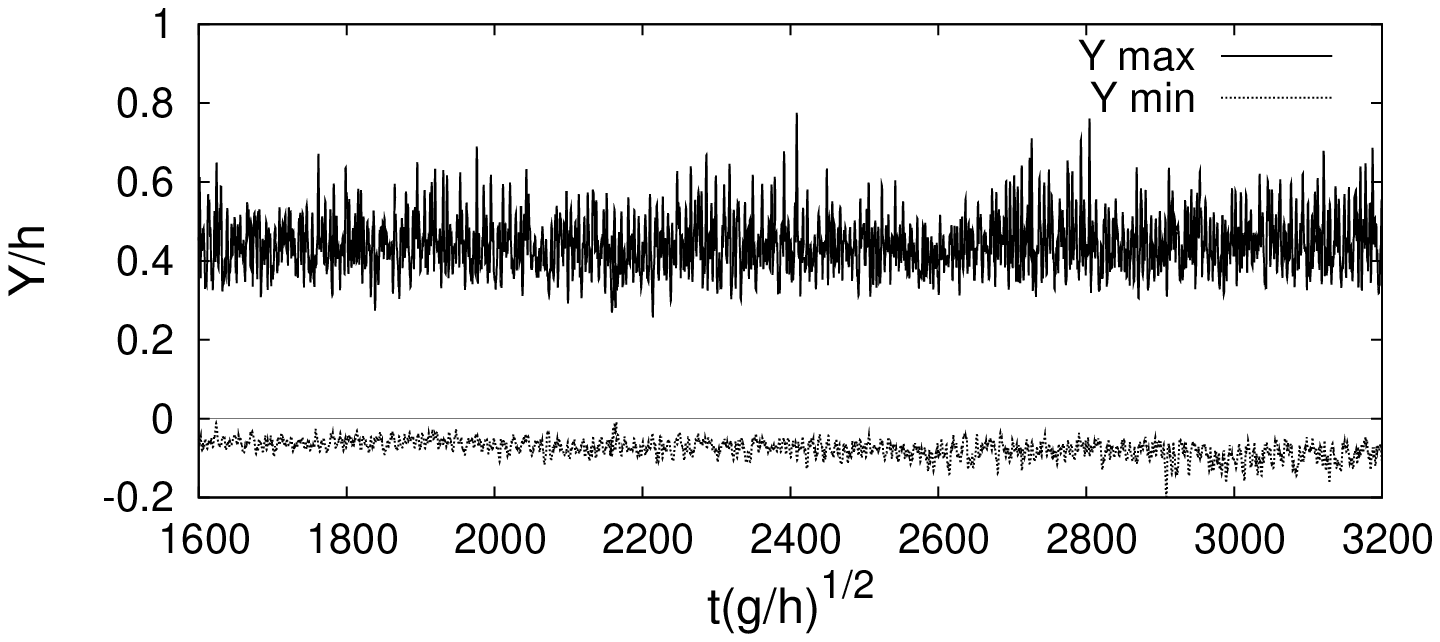,width=85mm}\\
   \epsfig{file=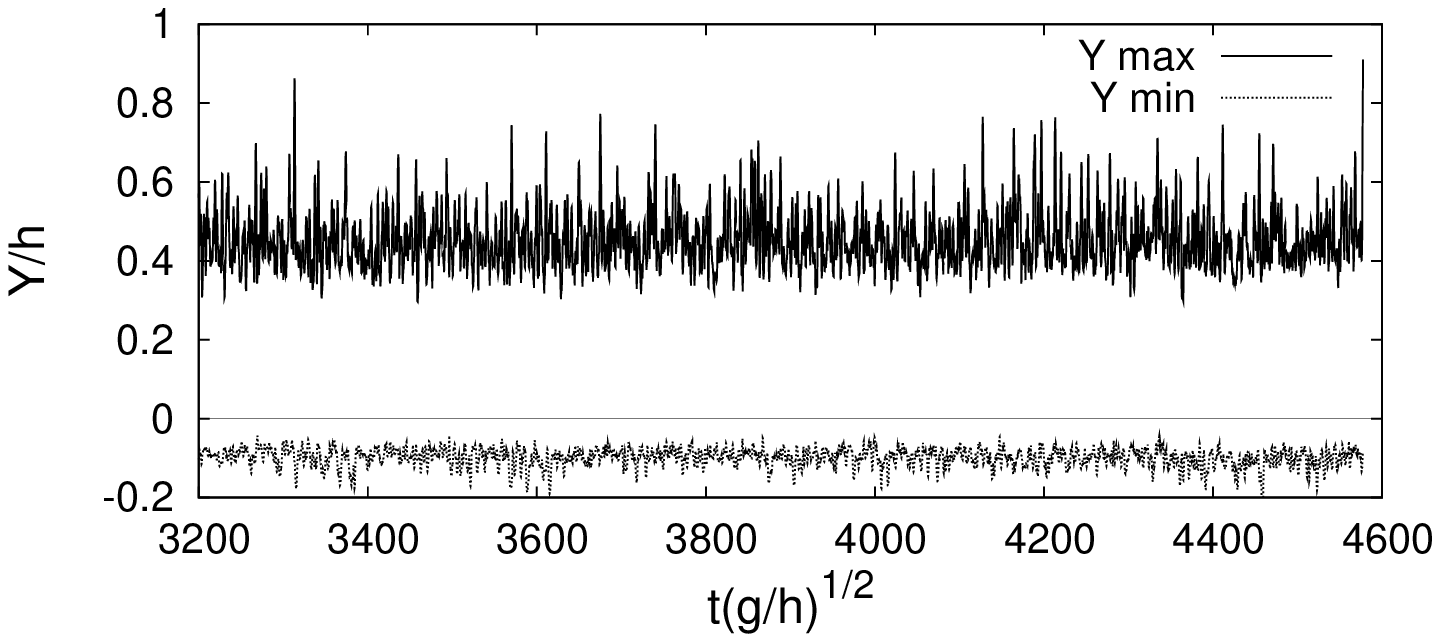,width=85mm}
\end{center}
\caption{Maximum and minimum elevations of the free surface over the mild-slope bed.} 
\label{smooth_bed_Ymax} 
\end{figure}

\begin{figure}
\begin{center}
   \epsfig{file=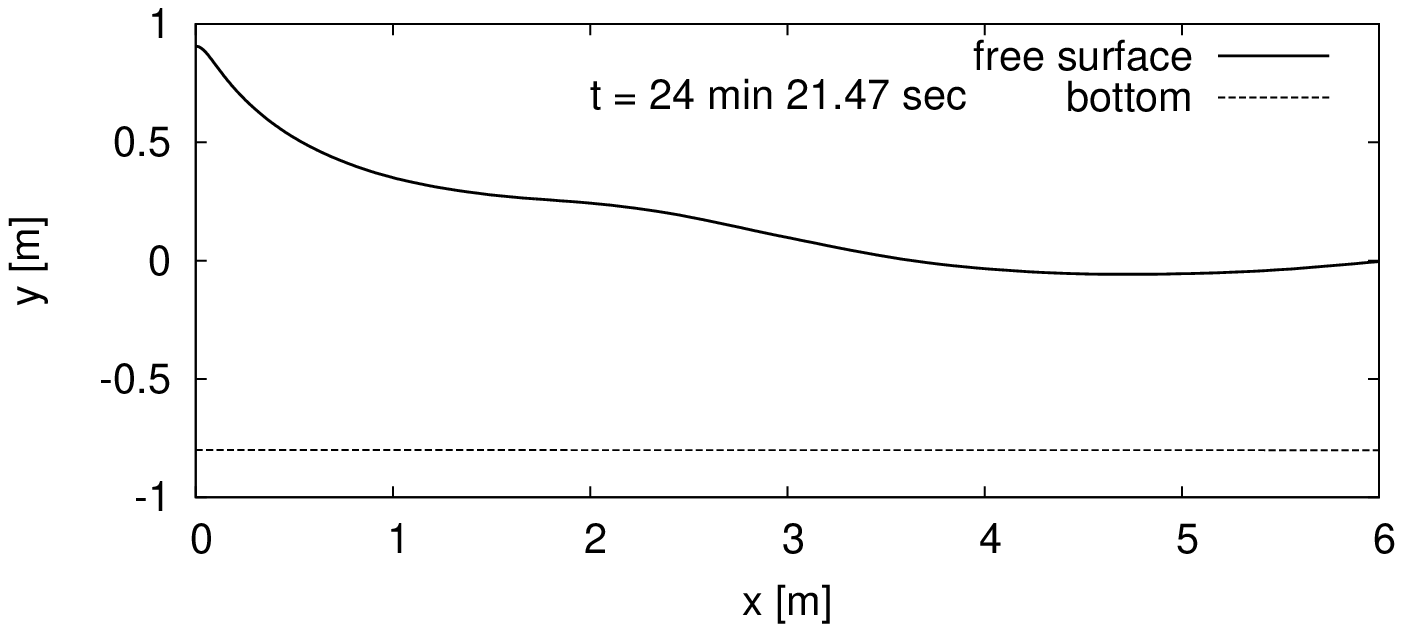,width=85mm}\\  
   \epsfig{file=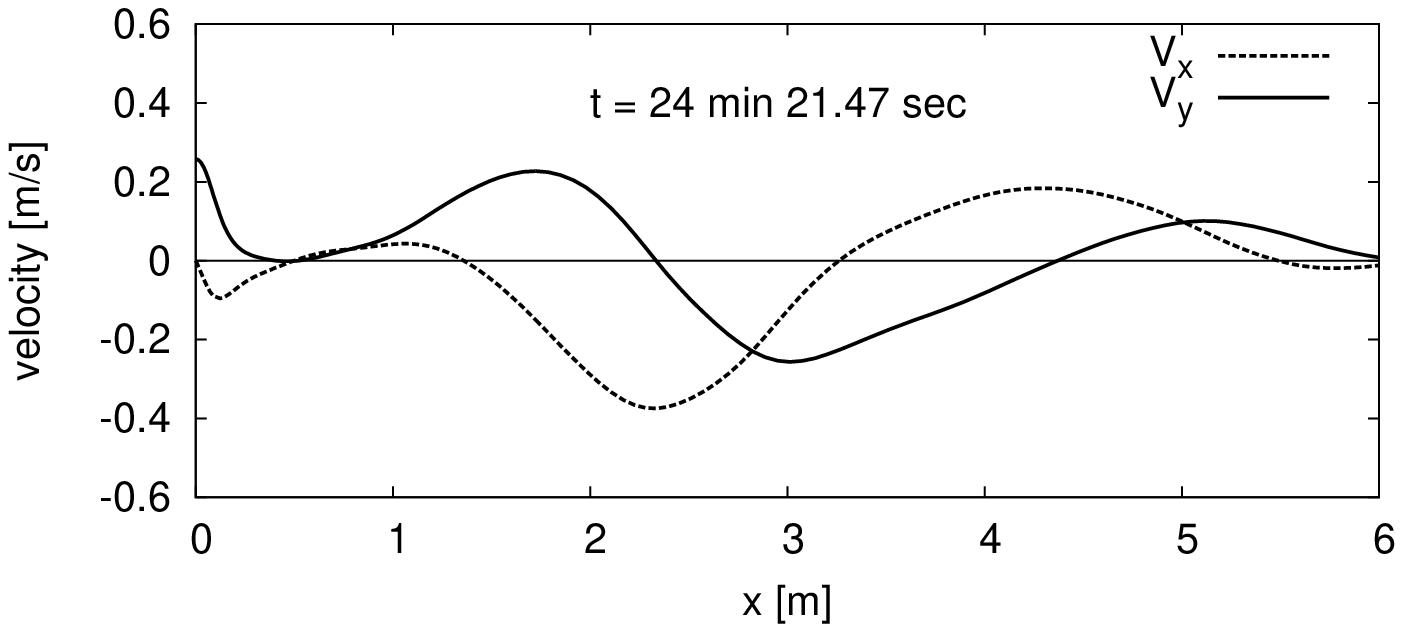,width=85mm}\\
\end{center}
\caption{Big wave over the mild-slope bed near the wall: a) the shape of free surface,
b) velocity distribution along the free surface.} 
\label{smooth_bed_wall_collision} 
\end{figure}

\begin{figure}
\begin{center}
   \epsfig{file=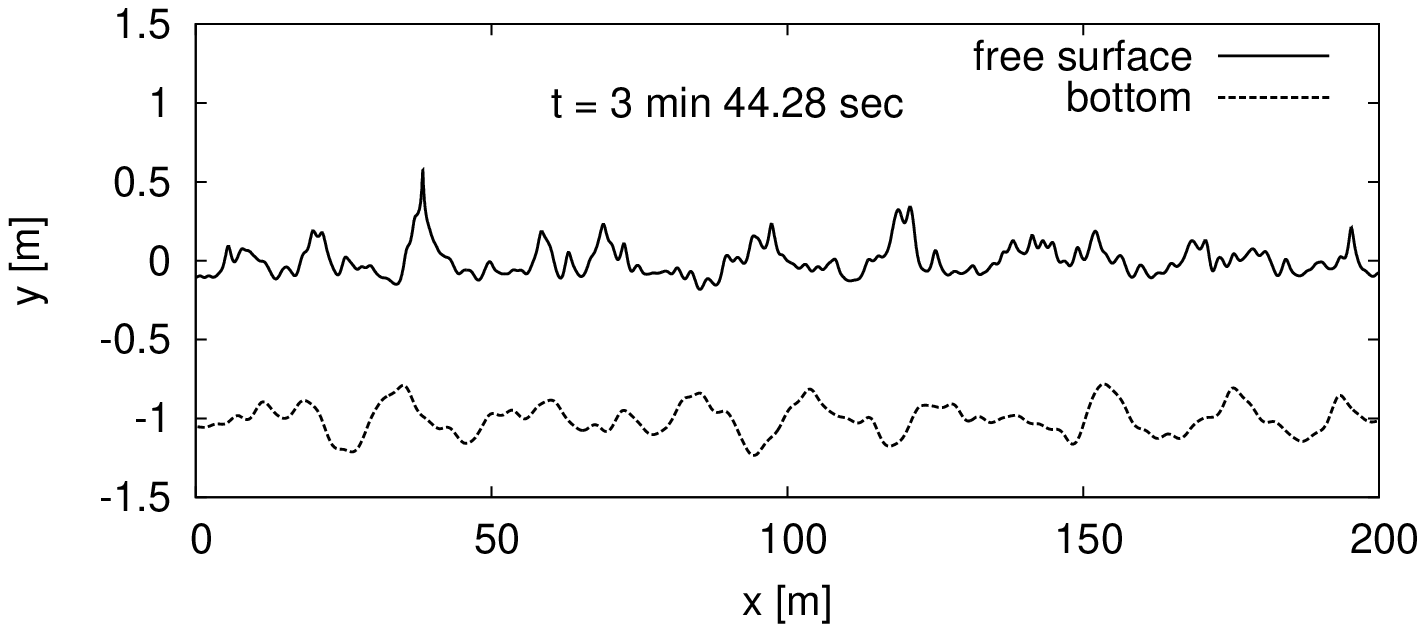,width=85mm}\\  
   \epsfig{file=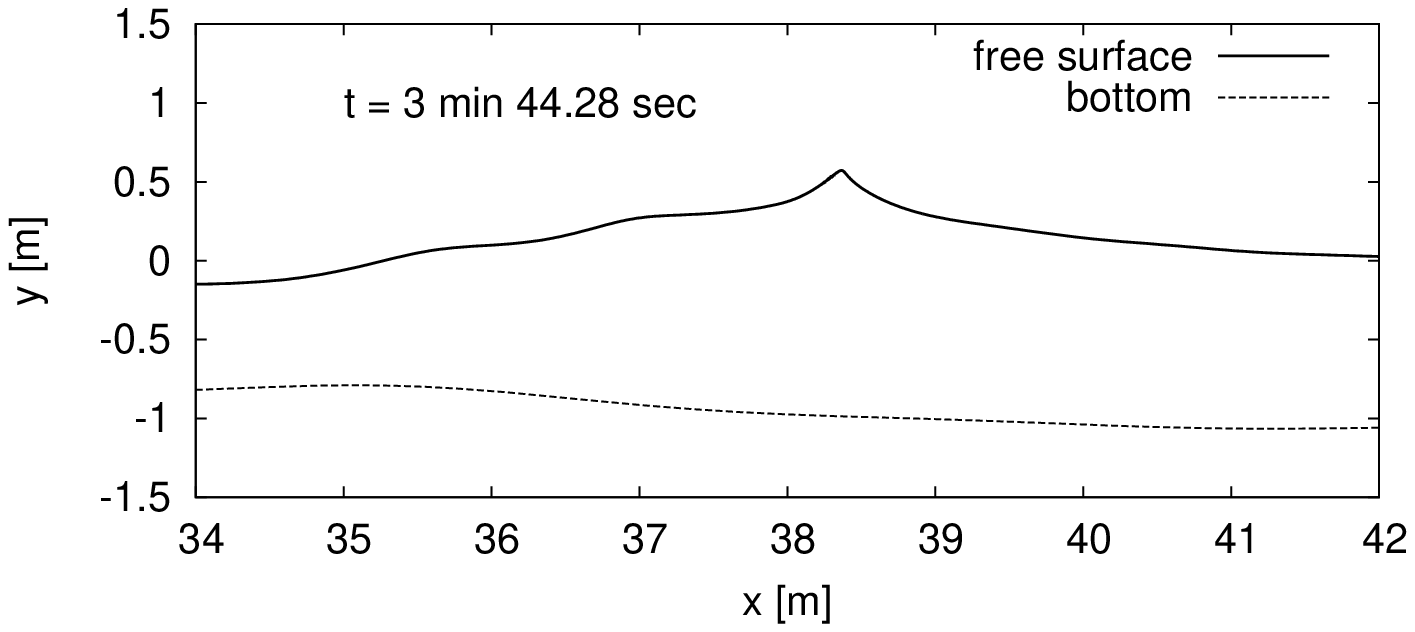,width=85mm}\\
   \epsfig{file=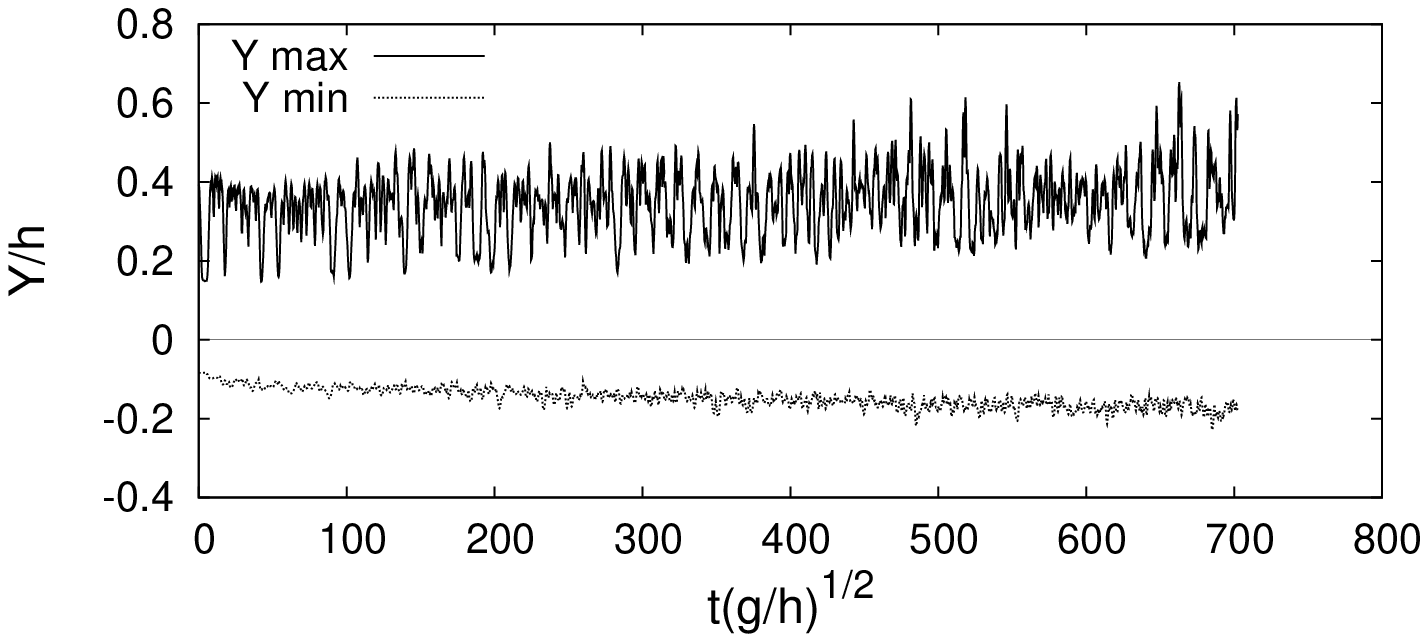,width=85mm}
\end{center}
\caption{Formation of sharp wave over a quasi-random bed.} 
\label{random_bed} 
\end{figure}

\begin{figure}
\begin{center}
   \epsfig{file=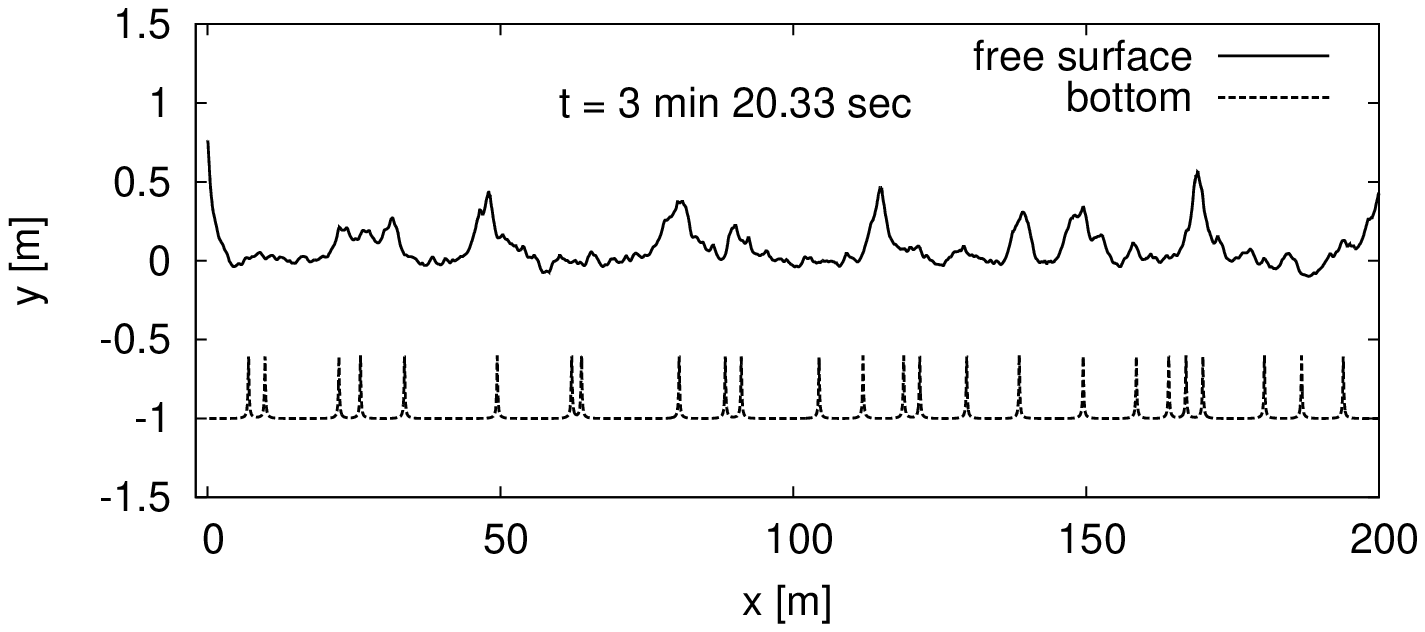,width=85mm}\\
   \epsfig{file=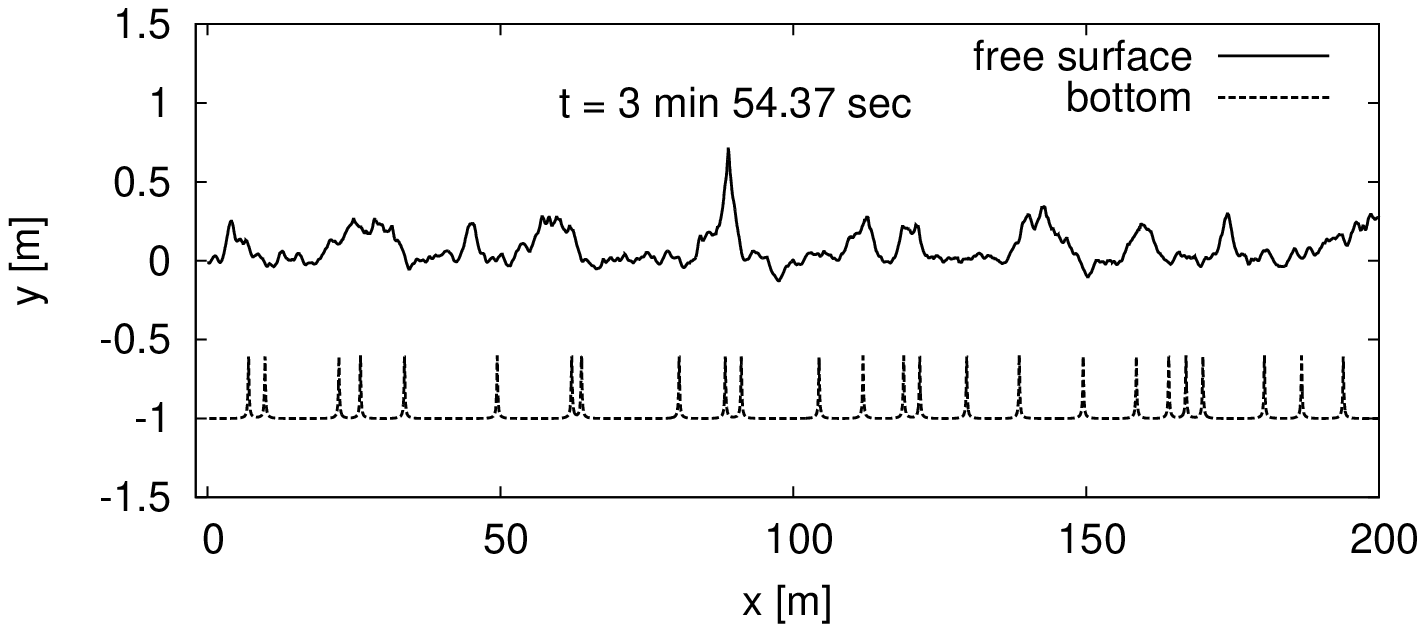,width=85mm}\\  
   \epsfig{file=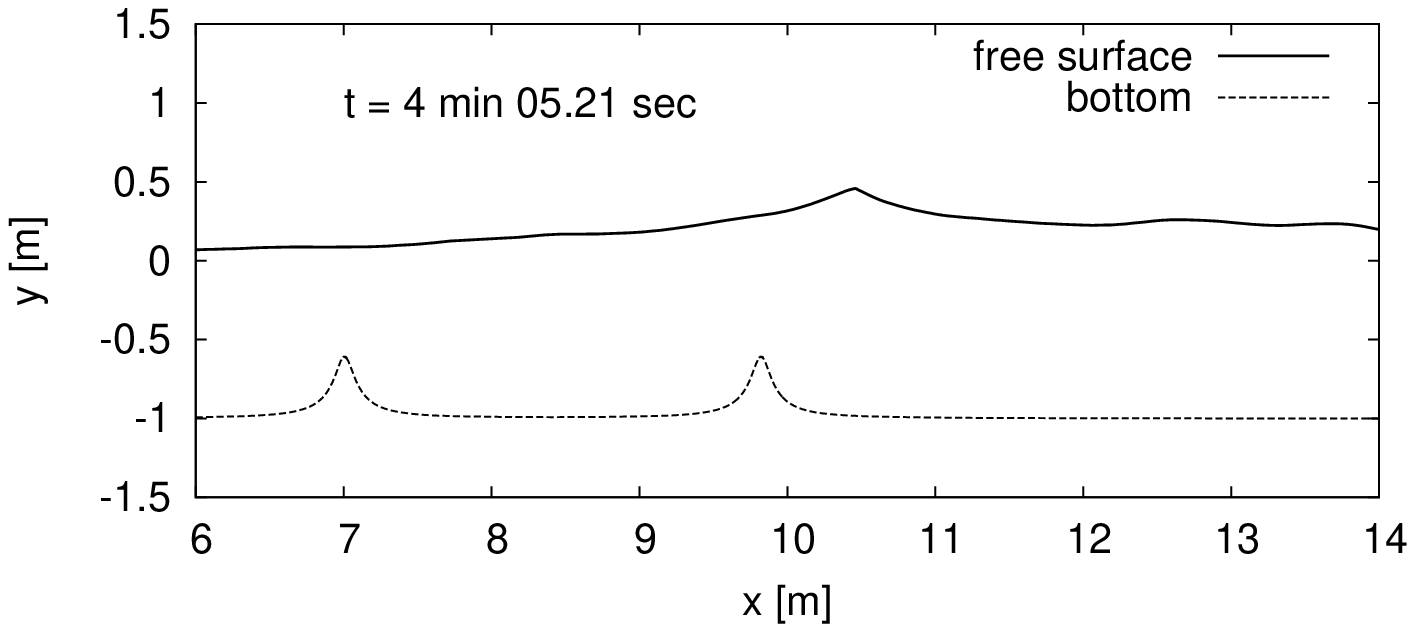,width=85mm}\\
   \epsfig{file=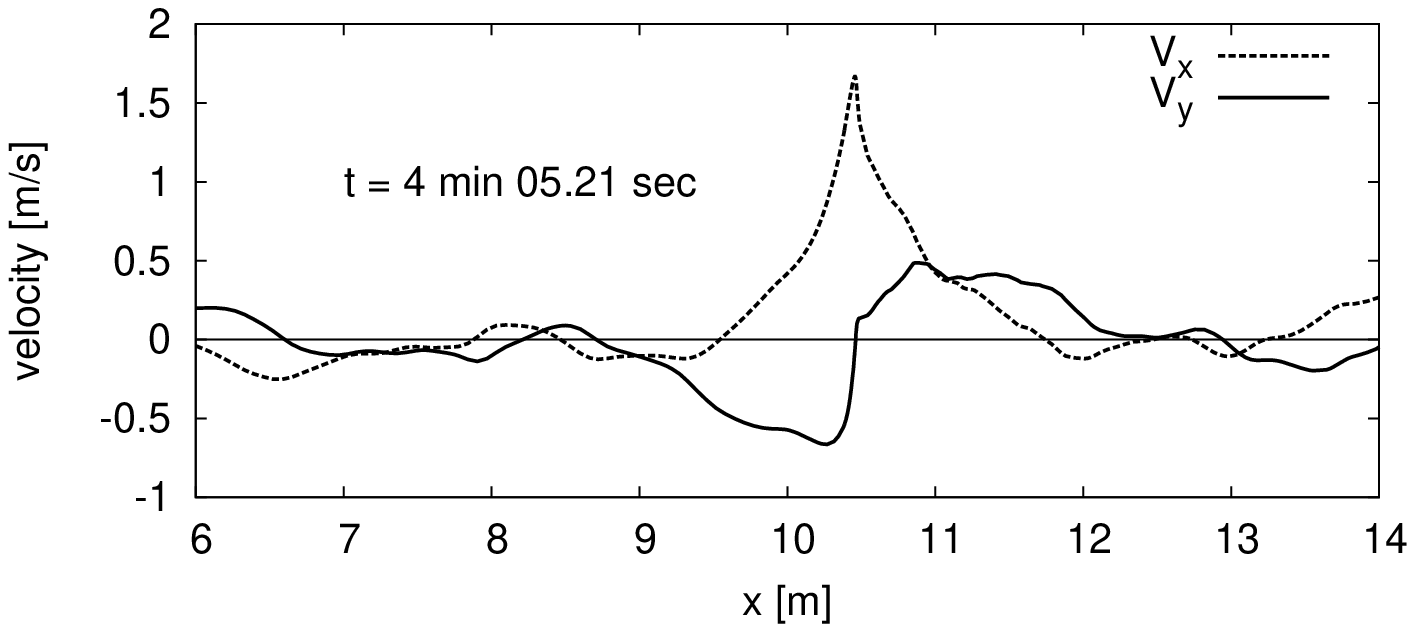,width=85mm}\\
   \epsfig{file=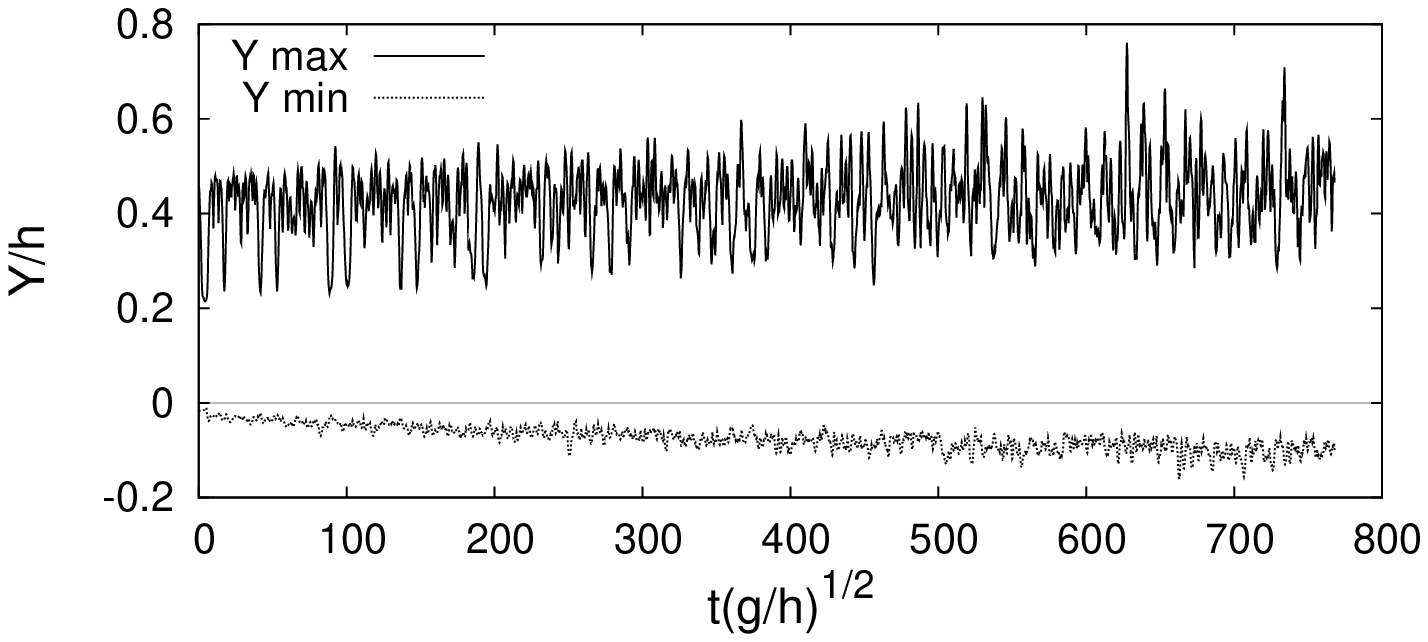,width=85mm}
\end{center}
\caption{Formation of extreme waves over a bed with randomly placed barriers.} 
\label{barriers_on_bed} 
\end{figure}

\section{Extreme waves over nonuniform beds}

We have observed through numerical experiments that with the flat bottom, 
if all the initial humps are nearly of the same height, then the maximum surface
elevation, as a function of time, does not exceed much the initial value for a long 
time (not shown). What happens if the bed is nonuniform? This question is answered 
in this Section through a set of numerical experiments. Three typical
examples are presented below.

In the first nonuniform case, we simulated waves over a mild-slope bottom, 
corresponding to function $Z(\zeta)$ in the form
\begin{equation}
Z_1(\zeta)=\zeta+i(2\pi/400)[0.2\exp(i\zeta)-1] 
\end{equation}
(see Fig.\ref{smooth_bed_big_waves}).
An initial state was eight pairs of colliding solitons placed at quasi-random positions 
[that is, eight humps of the form Eq.(\ref{collision})], with $s=1.11$ (not shown). 
Over nonuniform bed, the initial solitons evolve to a random wave field, consisting of
quasi-solitonic coherent structures of different heights, together with non-coherent
waves. After sufficiently long time, the smooth nonuniformity resulted in appearance of 
tall extreme wave events  
(see Figs.\ref{smooth_bed_big_waves}-\ref{smooth_bed_wall_collision}), 
when the most strong, oppositely propagating quasi-solitonic coherent structures collided. 
The highest wave events were observed near the left wall, where the depth is minimal.

In the second nonuniform case, we took function $Z(\zeta)$ as a sum of several
Fourier harmonics,
\begin{equation}
Z_2(\zeta)=\zeta-\frac{2\pi i}{400} +i \sum_{m=1}^8 (-1)^m C_m\exp(ik_m\zeta),
\end{equation}
with some positive coefficients $C_m$ and positive integer wave numbers $k_m$. 
Practically, in this case the bed profile looks as a quasi-random, relatively
short-correlated curve (see Fig.\ref{random_bed}).  
Again extreme waves appeared on a later stage of evolution, after the initial state 
consisting of eight pairs of colliding solitons with parameters $s=1.12$. 
Comparatively to the mild slope, now extreme waves were no so high, but very sharp
and typically more asymmetric. 
It should be also noted that in this case a typical time of the transition to a 
random wave field, characterized by a rough profile of the free surface, 
was essentially shorter than for the mild-slope bottom. Also, quasi-solitonic coherent
structures in the second case were relatively short-lived, while a non-coherent part 
of the wave filed was more developed.

In the third nonuniform case, the bed inhomogeneity was taken in the form of 25
randomly placed nearly identical barriers (shape of barriers is seen
in Fig.\ref{barriers_on_bed}), 
\begin{eqnarray}
&&Z_3(\zeta)=\zeta-\frac{2\pi i}{400}\nonumber\\
&&+\sum_{n}\Big[i\sqrt{\nu\exp[-B(\zeta+i\varepsilon-\zeta_n)^2]
-(\zeta+i\varepsilon-\zeta_n)^2}\nonumber\\
&&\qquad\qquad\qquad-(\zeta+i\varepsilon-\zeta_n)\Big],
\end{eqnarray}
where $\zeta_n$ for $n=0...24$ were quasi-random real numbers in the range 
from 0 to $\pi$, while $\zeta_{n+25}=2\pi-\zeta_n$ (required for the symmetry).
The other parameters were $B=1400$, $\varepsilon=0.001$, $\nu=5.0\times 10^{-5}$.
Contrary to the previous two examples, the conformal mapping, corresponding to
function $Z_3(\zeta)$, has singularities in the upper half-plane of the complex 
variable $\zeta$; however, all the singularities are far enough above the free surface.
The large value of the parameter $B$ allowed us to take into account only several nearest
barriers when the function $Z(\zeta)$ was numerically evaluated, thus significantly
reducing the computational cost comparatively to many different possible choices for
barrier shapes.

Eight initial pairs of solitons had $s=1.11$. In this numerical experiment, 
extreme waves appeared as well (see Fig.\ref{barriers_on_bed}). Some of them were 
quite tall, while other had a moderate height, but a very sharp crest (these were 
essentially asymmetric). In general, the third case is much similar to the second case 
(compare Fig.\ref{barriers_on_bed} and Fig.\ref{random_bed}).

\section{Summary and discussion}

In this work, it has been demonstrated through highly-accurate numerical simulations
of exact equations of motion for planar potential flows of a perfect fluid 
with a free surface, that shallow-water dispersive waves with moderate amplitudes 
$A/h\lesssim 0.12$ can exhibit the Fermi-Pasta-Ulam recurrence in a finite basin for
various initial states. However, the best quality of the recurrence
is observed for the initial free surface in the form $\eta_0(x)=A_0\cos(2\pi x/L)$,
and $\psi_0(x)=0$. In that special case, velocities of all arising solitons appear
self-consistently tuned to their phase shifts in mutual collisions, which property 
results in remarkably perfect recurrence to the initial state even in quite long flumes. 
A mathematical reason for such self-consistency is not clear at the moment.
The FPU quasi-recurrence is also robust with initial states in the form of two solitons.
For larger number of solitons, quasi-recurrence is possible with special values of
parameters only. 

All our numerical results are based on the inviscid theory. Of course, in the reality a 
viscous friction will act against the recurrence. However, in \cite{R2011Pisma} it was
estimated that a relative effect of the viscous friction near the bottom and near the side 
walls of the flume becomes small if all the spatial scales increase proportionally. 

In the quasi-integrable regime over a flat horizontal bed, with initial states in the form
of several nearly equal solitons, formation of extreme waves appears effectively suppressed, 
since the solitons preserve their strengths for a long time. 
When the approximate integrability destroyed by a bed nonuniformity, the system evolves
to a random-wave-field regime where quasi-solitonic coherent structures of different
amplitudes are present, some of them being stronger than the initial solitons. 
When the strongest oppositely propagating structures collide, fairly extreme waves arise. 
The most high extreme waves were observed for a mild-slope bed profile, while for
relatively short-correlated bed inhomogeneities the extreme waves were typically less tall 
but more sharp-crested. Similar effects were observed both for waves between 
the vertical walls, and for waves with periodic boundary conditions without the 
additional symmetry (not shown).

Our present results for extreme events in bidirectional wave fields over nonuniform 
beds may have some relevance to the problem of rogue (freak) waves in coastal zone, 
but only if the coast is in the form of a wave-reflecting cliff rather
than a wave-absorbing beach.

These investigations were supported by the Russian Foundation for
Basic Research (project no. 09-01-00631),
by the Council of the President of the Russian Federation for Support
of Young Scientists and Leading Scientific Schools (project no. NSh-6885.2010.2),
and by the Presidium of the Russian Academy of Sciences (program 
``Fundamental Problems of Nonlinear Dynamics'').

\appendix*

\section{A single soliton between the walls in the Boussinesq model}

The Boussinesq equations for weakly nonlinear, weakly dispersive long water waves
in the dimensionless variables [$\eta/h \to \eta$, $\sqrt{3}x/2h\to x$,
$\sqrt{3g/h}\,t/2\to t$] take the form
\begin{eqnarray}
&&u_t +uu_x+\eta_x=0,\\
&&\eta_t +[(1+\eta)u]_x+\frac{1}{4}u_{xxx}=0,
\end{eqnarray}
where $\eta$ is the vertical displacement of the free surface, and $u=\psi_x$
is a horizontal velocity. Following Zhang and Li \cite{ZhangLi2003}, we transform the 
above system to a more symmetric form
\begin{eqnarray}
\label{q_eq}
q_t+\frac{1}{2}q_{xx}+q^2r=0,&&\\
\label{r_eq}
-r_t+\frac{1}{2}r_{xx}+r^2q=0,&&
\end{eqnarray}
where new real unknown functions $q(x,t)$ and $r(x,t)$ express the old functions
$\eta(x,t)$ and $u(x,t)$ in the following manner:
\begin{equation}
\label{transform}
u=q_x/q,\qquad \eta=-1+qr+u_x/2.
\end{equation}
We note the system of equations (\ref{q_eq}) and (\ref{r_eq}) is formally similar
to the focusing nonlinear Shroedinger equation $2i\psi_t+\psi_{xx}+2\psi^2\psi^*=0$
and its complex conjugate $-2i\psi^*_t+\psi^*_{xx}+2{\psi^*}^2\psi=0$.
Therefore we can apply a simple generalization of the  
Akhmediev-Eleonskii-Korneev-Kulagin ansatz \cite{AK1986,AEK1987,AA1993}, 
and search for a solution of Eqs.(\ref{q_eq})-(\ref{r_eq})
in the form $(q,r)=[U(x,t)\pm \sqrt{Z(t)}]\exp(\pm P(t))$. 
Indeed, by doing so one can obtain and integrate a system of equations 
for the unknown functions $U(x,t)$, $Z(t)$, and $P(t)$. 
At some point, the problem is reduced to the analysis of two equations
(compare to \cite{AEK1987}):
\begin{equation}
\dot Z^2-16Z^4 +16 w Z^3 -4(h+w^2)Z^2 -4bZ=0,
\end{equation}
\begin{equation}
U_x^2+U^4 +2(w-3Z)U^2 +2\frac{\dot Z}{\sqrt{Z}} U +(2wZ-3Z^2-b)=0,
\end{equation}
where $w,h,b$ are some constants (there is also the third equation $\dot P+2Z = w$).

However, we prefer not to deal with a function of two variables as $U(x,t)$, 
and therefore we use here  a slightly less general ansatz which still contains 
physically interesting solutions, and where the variables are separated from 
the very beginning:
\begin{eqnarray}
\label{FQA}
q(x,t)&=&F(t)+\frac{Q(t)}{D(x)+A(t)},\\ 
\label{GRA}
r(x,t)&=&G(t)+\frac{R(t)}{D(x)+A(t)}.
\end{eqnarray}
We take the only $x$-dependent function $D(x)$ satisfying the relations
\begin{eqnarray}
\label{D_xD_x}
D_x^2&=&4\mu^2(D^2-1)(1-\epsilon^2D^2)\nonumber\\
&\equiv& cD^2-\delta D^4 -\beta,\\
\label{D_xx}
D_{xx}&=&cD-2\delta D^3.
\end{eqnarray}
Thus, it is one of the elliptic Jacobi functions (for their definitions 
and properties, see, e.g., \cite{WW}):
\begin{equation}
D(x)=\mbox{nd}(2\mu x,\sqrt{1-\epsilon^2}).
\end{equation}
The $x$-period of this function is $\tilde L=I(\epsilon)/\mu$, where
\begin{equation}
I(\epsilon)=\int_1^{1/\epsilon}\frac{dz}{\sqrt{(z^2-1)(1-\epsilon^2z^2)}}.
\end{equation}

Now we substitute the ansatz (\ref{FQA})-(\ref{GRA}) into the system
(\ref{q_eq})-(\ref{r_eq}). Using relations (\ref{D_xD_x}) and (\ref{D_xx}), 
we obtain the following set of equations (which are coefficients in front of 
different powers $(D+A)^{-n}$, for $n=0,1,2,3$, or their linear combinations):
\begin{eqnarray}
\dot F +F^2G+\delta A Q=0,&&\\
-\dot G +FG^2+\delta A R=0,&&\\
\dot Q +\frac{Q}{2}(c-6\delta A^2) +2QFG+RF^2=0,&&\\
-\dot R +\frac{R}{2}(c-6\delta A^2) +2RFG+QG^2=0,&&\\
2\dot A+QG-RF=0,&&\\
-(cA-2\delta A^3)+QG +RF=0,&&\\
(cA^2-\delta A^4-\beta) +QR=0.
\end{eqnarray}
(The last equation actually appears twice.)
It is easy to show that the two algebraic relations are consistent with the 
five differential equations. It follows from these equations also that 
\begin{equation}
FG=\delta A^2+\gamma,
\end{equation}
where $\gamma$ is a constant. Now we take the squared equation for $\dot A$ and obtain
$4\dot A^2=(RF-QG)^2=(RF+QG)^2-4FGQR=
(cA-2\delta A^3)^2 +4(\gamma+\delta A^2)(c A^2-\delta A^4 -\beta)$, 
that is an easily solvable first-order equation
\begin{eqnarray}\label{dotA}
4\dot A^2&=&-4\gamma\delta A^4 +A^2(c^2+4c\gamma-4\beta\delta)-4\gamma \beta\nonumber\\
&\equiv&4\gamma\delta(A^2-\alpha_1^2)(\alpha_2^2-A^2),
\end{eqnarray}
where
\begin{equation}
\alpha^2_{1,2}=\frac{1}{2}\left[
\frac{c^2}{4\gamma\delta}+\frac{c}{\delta}-\frac{\beta}{\gamma}\right]
\mp
\sqrt{\frac{1}{4}\left[
\frac{c^2}{4\gamma\delta}+\frac{c}{\delta}-\frac{\beta}{\gamma}\right]^2
-\frac{\beta}{\delta}}.
\end{equation}
The solution of Eq.(\ref{dotA}) is again expressed through an elliptic function:
\begin{equation}
A(t)=\alpha_1\, \mbox{nd}\left( t\alpha_2\sqrt{\gamma\delta},
\sqrt{1-\left[\frac{\alpha_1}{\alpha_2}\right]^2}\right)
\equiv\alpha_1\, \mbox{nd}\left(\xi,\kappa\right).
\end{equation}
Since, by definition, $\mbox{nd}(\xi,\kappa)=1/\mbox{dn}(\xi,\kappa)$ (see \cite{WW}), 
and $[\mbox{dn}(\xi,\kappa)]_\xi=-\kappa^2\mbox{sn}(\xi,\kappa)\,\mbox{cn}(\xi,\kappa)$, 
for the time derivative $\dot A(t)$ we have
\begin{equation}
\dot A(t)=\alpha_1 \alpha_2\sqrt{\gamma\delta}\kappa^2
\mbox{sd}(\xi,\kappa)\,\mbox{cd}(\xi,\kappa).
\end{equation}

Thus, we have obtained explicit expressions for the quantities $A$, $QR$, $FG$, and
$(RF+QG)$. From these one can also extract the relations $Q/F$ and $R/G$, since
\begin{eqnarray}
&&\frac{R}{G}+\frac{Q}{F}=\frac{(RF+QG)}{FG}=\frac{(cA-2\delta A^3)}{\delta A^2+\gamma},\\
&&\frac{R}{G}-\frac{Q}{F}=\frac{2\dot A}{FG}=\frac{2\dot A}{\delta A^2+\gamma}.
\end{eqnarray}
The obtained information is sufficient to construct the velocity $u(x,t)$ and
the free surface elevation $\eta(x,t)$ via formulas (\ref{transform}), because
\begin{eqnarray}
u&=&\left[\frac{1}{(D+A+Q/F)}-\frac{1}{(D+A)}\right]D_x,\\
\eta&=&-1+FG+\frac{(RF+QG)}{(D+A)}+\frac{QR}{(D+A)^2} \nonumber\\
&& +u_x/2.
\end{eqnarray}
Using relations $c=4\mu^2(1+\epsilon^2)$, $\delta=4\mu^2\epsilon^2$, 
and $\beta=4\mu^2$, it is possible to show that $\alpha_1<1$ and $\alpha_2>1/\epsilon$, 
and therefore at definite time moments function $A(t)$ takes values $A_1=1$ 
or $A_2=1/\epsilon$. Simultaneously, at those time moments  either $Q=0$, or $R=0$. 
When $Q=0$, then the velocity field $u(x)$ is zero everywhere, 
while the free surface profile is either $\eta_1(x)$, or $\eta_2(x)$, where
\begin{eqnarray}
\eta_1(x)&=&-1+\gamma +4\mu^2\left[\epsilon^2 +\frac{(1-\epsilon^2)}{D(x)+1}\right],\\
\eta_2(x)&=&\eta_1(x-\tilde L/2),
\end{eqnarray}
 We see that the best choice for the constant
$\gamma$ is $\gamma=1-4\mu^2 \epsilon$, since in this case $\eta_{1 \rm min}=0$.
Function $\eta_1(x)$ has a single hump at $x=0$, and thus it
corresponds to the moments when a soliton is colliding with the left
wall, while $\eta_2(x)$ corresponds to the collisions of the soliton 
with the right wall at $x=\tilde L/2$.  

In the limit $\epsilon \ll 1$ we have $\tilde L\to\infty$,
and $D(x)\approx \cosh(2\mu x)$, so  
\begin{equation}
\eta_1(x)\approx\frac{4\mu^2}{1+\cosh(2\mu x)}=\frac{2\mu^2}{\cosh^2(\mu x)}.
\end{equation}
The full solution in this limit is given by the following formulas
(it is interesting to note the solution below is essentially Eq.(52) of Ref.\cite{AEK1987} 
for the focusing nonlinear Schroedinger equation, but evaluated at imaginary time):
\begin{widetext}
\begin{eqnarray}
q(x,t)&=&\left[1+\frac{2\mu^2 \cosh(2t\mu\sqrt{1+\mu^2})
-2\mu\sqrt{1+\mu^2} \sinh(2t\mu\sqrt{1+\mu^2})}
{\sqrt{1+\mu^2}\cosh(2\mu x)
+\cosh(2t\mu\sqrt{1+\mu^2})}\right]\mbox{e}^{-t},\\
r(x,t)&=&\left[1+\frac{2\mu^2 \cosh(2t\mu\sqrt{1+\mu^2})
+2\mu\sqrt{1+\mu^2} \sinh(2t\mu\sqrt{1+\mu^2})}
{\sqrt{1+\mu^2}\cosh(2\mu x)
+\cosh(2t\mu\sqrt{1+\mu^2})}\right]\mbox{e}^{t}
\end{eqnarray}
\end{widetext}
Collision of the soliton with the wall at $x=0$ occurs at $t=t_*$, when
$$
\cosh(2t_*\mu\sqrt{1+\mu^2})=\sqrt{1+\mu^2}.
$$
It is easy to derive that before and after the collision, the soliton 
(at $x> 0$) moves with the velocities $s=\mp\sqrt{1+\mu^2}$, respectively.

\end{document}